\documentclass[%
twocolumn,
superscriptaddress,
 amsmath,amssymb,
 prl,
]{revtex4-1}
\usepackage{lineno}

\usepackage{xcolor}
\usepackage{graphicx}
\usepackage{dcolumn}
\usepackage{bm}
\usepackage{siunitx}
\DeclareSIUnit\gauss{G}
\usepackage{comment}
\newcommand{\todiscuss}[1]{\textcolor{black}{#1}}
\newcommand{\reviewtwo}[1]{\textcolor{black}{#1}}
\newcommand{\reviewthree}[1]{\textcolor{black}{#1}}

\begin{document}

\preprint{}

\title{\reviewtwo{Phase-Space Shaping due to Betatron Cooling in Wakefield Accelerators\\ and the Onset of the Ion-Channel Laser}}

\author{P. J. Bilbao}%
 \email{pablo.bilbao@lmh.ox.ac.uk}
\affiliation{%
 GoLP/Instituto de Plasmas e Fus\~{a}o Nuclear, Instituto Superior T\'{e}cnico, Universidade de Lisboa, 1049-001 Lisbon, Portugal
}%
\affiliation{The Rudolf Peierls Centre for Theoretical Physics, University of Oxford, Oxford OX1 3NP, UK}

\author{T. Silva}
\author{L. O. Silva}%
 \email{luis.silva@tecnico.ulisboa.pt}
\affiliation{%
 GoLP/Instituto de Plasmas e Fus\~{a}o Nuclear, Instituto Superior T\'{e}cnico, Universidade de Lisboa, 1049-001 Lisbon, Portugal
}%

\date{\today}

\begin{abstract}
Plasma-based accelerators are beginning to employ relativistic beams with unprecedented charge and ultrashort durations. These dense driver beams can drive wakes even in high-density plasmas ($\gtrsim10^{19}$ cm$^{-3}$), where betatron radiation becomes increasingly important and begins to affect the dynamics of the accelerated beam. We show that betatron cooling leads to a strong structuring of the phase space of the beam. This gives rise to bunched, ring-like structures with positive radial position and momentum gradients, \emph{i.e.}, population inversion of the amplitude of oscillation. We derive the characteristic timescales for this process analytically and confirm our predictions with multi-dimensional Particle-in-Cell simulations. The radiation-dominated regime of beam dynamics fundamentally alters the acceleration process and produces self-structured beams\reviewtwo{; Particle-in-Cell simulations of the resulting beams show that it becomes unstable to the ion-channel laser and lead to exponential amplification of radiation.}
\end{abstract}

\maketitle

\reviewtwo{Plasma accelerators sustain gradients up to $100\, \mathrm{GV}/\mathrm{m}$ \cite{tajima_laser_1979,chen1985acceleration,wang2002x,mangles2004monoenergetic,faure2004laser,geddes2004high,hogan2005multi,adli2018acceleration,gonsalves2019petawatt}, enabling multi-GeV energy gains \cite{hogan2005multi,gonsalves2019petawatt}. Pushing toward tighter, higher-charge driver beams and denser plasmas yields even greater gradients \cite{joshi2018plasma, joshi2020perspectives}. accompanying intense focusing fields cause off-axis accelerating electrons to undergo transverse betatron oscillations \cite{pukhov2002strong}, a well-known source of intense radiation in laser- and beam-driven schemes \cite{whittum1990ion, chen1991unified,whittum1992electromagnetic, wang2002x, esarey2002synchrotron,pukhov2002strong, clayton2009towards, kneip2010bright, corde2013femtosecond,arefiev2016beyond,vranic2018extremely,babjak2024direct}. }

\reviewtwo{In intense fields, relativistic electrons undergoing betatron oscillations radiate energy at rates comparable to their acceleration} \cite{rousse2004production,pukhov2002strong}. 
As plasma-based accelerators can now deliver high charge ($\mathcal{O}(q)\sim\ $nC), short driver beams ($\mathcal{O}(\sigma_z)\sim5\ \mu$m), these beams can even drive wakes in high-density plasmas ($\mathcal{O}(n_0)\gtrsim10^{19}\,\mathrm{cm}^{-3}$) operating in the blowout regime \cite{joshi2020perspectives}.
\reviewtwo{Centimetre-scale plasma sources, beam-driven waves at these densities, and multi-GeV wakefield beams have all been demonstrated experimentally \cite{aniculaesei2018novel,gilljohann2019direct,zhang2025plasma}.}

In such regimes, radiative losses become increasingly important and must be self-consistently incorporated into the kinetic description of the beam dynamics. Previous studies have addressed radiative effects such as emittance damping and energy loss \cite{lee1982radiation,michel2006radiative}. 
\todiscuss{Recent experiments have demonstrated that plasma wakefield accelerators can reproducibly generate high-quality, self-matched witness beams with energies exceeding 20~GeV and strongly enhanced brightness, with injection and acceleration occurring in high-density plasma regions. The level of experimental control now achievable in beam-driven wakefields supports access to the radiation-relevant regimes considered here \cite{zhang2025plasma}.}


\todiscuss{In this Letter, we go beyond existing single-particle and moment-based descriptions of betatron cooling and analyze the full kinetic evolution of the transverse phase-space distribution. We show that the intrinsic nonlinearity of radiation reaction with respect to the betatron oscillation amplitude leads to a fundamental change  of the distribution function and the emergence of phase-space structuring.} We show that betatron cooling, \emph{i.e.}, radiation reaction from transverse oscillations in ion channels, will strongly modify the phase-space dynamics of high-energy beams, going beyond traditional descriptions of emittance damping where this kinetic bunching is not fully described \cite{lee1982radiation}. The radiated power scales as $P \propto \gamma^4 \left(\mathbf{p}\times \mathbf{a}\right)^2$ \cite{lightman1982relativistic}, where $\mathbf{p}$ is the momentum of the particle, $\gamma$ is the Lorentz factor, $\mathbf{a}$ is the acceleration, and $\left|\mathbf{a}\right|\propto A$. This already demonstrates that electrons in different regions of phase space cool at different rates (much like in magnetic synchrotron radiation, where $|\mathbf{a}| \propto p_\perp$ \cite{bilbao2022radiation}).\todiscuss{The amplitude dependence of radiative damping has been identified previously \cite{michel2006radiative}, but its implications for the full kinetic evolution of the full transverse distribution function have not been addressed. Previous works did not identify that since the amplitude $A$} depends on the initial displacement from the channel axis $x$ and the transverse momentum $p_x$, \todiscuss{then} radiative cooling induces bunching in phase space. 
\todiscuss{We show that this behavior emerges generically for beams with energies  $E \gtrsim 10\,\mathrm{GeV}$ in dense plasmas $n_0 \gtrsim 10^{19}\,\mathrm{cm}^{-3}$, as a cumulative effect of radiation-reaction-induced damping rather than from fine-tuned initial conditions or phase-mixing alone.}

We analytically derive the relevant timescales and structural evolution, and validate our predictions with multi-dimensional particle-in-cell (PIC) simulations performed using OSIRIS \cite{fonseca2002osiris,davidson2015implementation, vranic2015particle, vranic2016classical_m, vranic2016quantum_m}. These structured beams exhibit properties that distinguish them from those produced in conventional laser wakefield acceleration (LWFA) \cite{tajima_laser_1979} or plasma wakefield acceleration (PWFA) schemes \cite{chen1985acceleration, wang2002x}. Their sharp gradients in momentum space and ring-like profiles in transverse phase space make them ideal candidates to drive the ion-channel laser (ICL) mechanism \cite{whittum1990ion,esarey2002synchrotron,davoine2018ion}. Moreover, these beams develop naturally from the radiative dynamics of beam-plasma interaction, offering a new pathway to observe phase-space bunching effects analogous to those predicted for synchrotron radiation in magnetic systems \cite{bilbao2022radiation,bilbao2024ring,bilbao2024radiative}.

We consider the dynamics of an electron beam with Lorentz factor $\gamma\gg1$ propagating through a plasma of background density $n_0$ and the electron plasma frequency $\omega_{p} = (4\pi n_0 e^2 / m_e)^{1/2}$, where $e$ is the elementary charge and $m_e$ the electron mass. The beam travels within a blowout, or bubble, structure excited by either a beam or laser driver \cite{lu2006nonlinear}. The process described here is independent of the specific driver, making the results applicable to PWFA, LWFA, and ion-channel scenarios alike \cite{ng2001observation, hogan2003ultrarelativistic}. We describe the motion of electrons in the full blowout fields; each electron trajectory can be simplified to an oscillation in a plane determined by its initial position and momentum. For this reason we describe the motion relative to the blowout center and assume the transverse motion is restricted to the $x$ direction in the plane, \textit{i.e.}, $\mathbf{r} = (x, \xi)$ and $\mathbf{p} = (p_x, p_z)$, where $\xi = z - v_\mathrm{blowout} t -z_0$ is the co-moving longitudinal coordinate relative to the center of the blowout structure $z_0$. The blowout electric and magnetic fields inside the cavity are given by $E_z = \frac{1}{2} (m_e \omega_p^2 / e)\, \xi$, $E_x = \frac{1}{4} (m_e \omega_p^2 / e)\, x$, and $B_y = -\frac{1}{4} (m_e \omega_p^2 / e)\, x$, and the Lorentz force on a relativistic electron is $\mathbf{F}_\mathrm{L} = -\frac{1}{2} m_e \omega_p^2 \mathbf{r}$ \cite{lu2006nonlinear,esarey2009physics, vieira2016multidimensional}.

The particles in the beam experience radiative losses due to their strong transverse acceleration. For an electron in a wakefield accelerator the quantum parameter $\chi$ scales as $\chi\simeq 0.1 \frac{n_0}{\SI{e20}{\centi\metre^{-3}}} \frac{A}{\SI{10}{\micro\metre}} \frac{E}{\SI{10}{\giga\electronvolt}}$, where $\chi$ is a Lorentz- and gauge-invariant quantity defined as $\chi = e\sqrt{-(F_{\mu\nu} p^\nu)^2} / (E_\mathrm{Sc} m_e^3)$, with $F_{\mu\nu}$ the electromagnetic tensor, $E_\mathrm{Sc}\simeq4.41\times10^{13}\,\mathrm{statV}/\mathrm{cm}$ the Schwinger critical field, and $p^\nu$ the particle four-momentum~\cite{ritus1985quantum, di2012extremely}. From this scaling, we find that present wakefield experiments operate within the classical regime of betatron radiation ($\chi \ll 1$). Under these conditions, the classical description of radiation reaction is adequate. In contrast, when $\chi \gtrsim 1$, photon emission becomes stochastic and induces diffusion in phase space while still permitting the formation of bunching structures~\cite{bilbao2022radiation, bilbao2024ring}. We therefore expect that the qualitative behavior identified here will persist in the transition region $\chi \sim \mathcal{O}(1)$.

\reviewtwo{The classical radiative force, to first order in $\gamma$, is described by the Landau–Lifshitz formulation} \cite{landau1975classical, blackburn2020radiation} (in c.g.s. units):
\begin{equation}
    \mathbf{F}_\mathrm{RR} = -\frac{2}{3} \frac{r_e^2}{ c} \frac{\gamma \mathbf{p}}{m_e c} \left[\left(\mathbf{E} + \frac{\mathbf{p}\times \mathbf{B}}{\gamma m_e c}\right)^2 -\left( \frac{\mathbf{p}\cdot\mathbf{E}}{\gamma m_e c}\right)^2\right],\label{eq:LL}
\end{equation}
where $\alpha$ is the fine-structure constant, $c$ is the speed of light, \todiscuss{$r_e=e^2/m_ec^2$} is the classical electron radius \todiscuss{in c.g.s.}, and $\mathbf{E}$ and $\mathbf{B}$ are the electric and magnetic field, respectively. 

\reviewtwo{For relativistic beams ($\gamma\gg1$, $p_z \gg p_x$) in the blowout regime, the bracketed terms in Eq.~\eqref{eq:LL} simplify to $\mathbf{F}_L^2 \simeq E_z^2 + (E_x-B_y)^2$ and} $(\mathbf{p}\cdot\mathbf{E}/\gamma m_e c)^2\simeq E_z^2$. As a result, the radiation reaction force depends only on the transverse displacement from the blowout axis $x$, \emph{i.e.}, $\mathbf{F}_\mathrm{RR} =-(r_e^2m_e^2\omega_p^2/6ce^2)(\gamma \mathbf{p}/m_ec)x^2$. 
Note that most of the radiative loss occurs in the parallel direction, since $\mathbf{F}_\mathrm{RR}\propto \mathbf{p}$ and $p_z \gg p_x$, and at the turns of the betatron motion, where $x^2$ is maximum. This is a key feature of betatron cooling: the cooling rate depends nonlinearly on the betatron oscillation amplitude $F_\mathrm{RRx}\propto p_x x^2\propto A^3$. Similarly to how synchrotron cooling depends nonlinearly on the particle momentum, leading to bunching of the distribution function in momentum \cite{bilbao2022radiation, bilbao2024radiative}, the non-linearity of betatron cooling will lead to bunching in phase space.  

\reviewtwo{The equations of motion, normalizing lengths to $c/\omega_{p}$, time to $\omega_{p}^{-1}$, mass to $m_e$, velocities to $c$, and charge squared to $e^2=r_e/m_e c^2$, are}
\begin{subequations}
\begin{align}
    \frac{d\mathbf{r}}{dt} &=  \frac{\mathbf{p} }{\gamma} ,\label{eq:dr}\\
    \frac{d p_z}{dt} &= -\frac{1}{2} \xi - \frac{r_e}{6}  \gamma\,p_z x^2,\label{eq:dp_para}\\
    \frac{dp_x}{dt} &= -\frac{1}{2} x -\frac{r_e}{6}   \gamma\,  p_x x^2,\label{eq:dp_perp}
\end{align}
\end{subequations}
where the first term of Eqs. \eqref{eq:dp_para} and \eqref{eq:dp_perp}, proportional to $\xi$ and $x$, correspond to the Lorentz force and the second term, proportional to $x^2$, arises from the radiation reaction force (Eq.~\eqref{eq:LL}). The electron undergoes damped harmonic oscillations, radiating both its perpendicular momentum (emittance damping) and its parallel momentum. 
\reviewtwo{Equation \eqref{eq:dp_para} identifies an equilibrium position $\xi_{\mathrm{eq}}$ where wakefield momentum gain balances radiative losses over a betatron period is} $\left<dp_z/dt\right>_{\beta} =0$ in Eq. \eqref{eq:dp_para}, where $\left<\cdot\right>_{\beta}$ represents the quantity averaged over a betatron period, leading to $\xi_\mathrm{eq} = - r_e \gamma\, p_z  \left<x^2\right>_\beta$/3. Electrons at $\xi < \xi_{\mathrm{eq}}$ can gain parallel momentum over each oscillation, whereas those at $\xi > \xi_{\mathrm{eq}}$ radiate more energy than they gain.  As a consequence, future high-density and high-energy wakefield accelerators must ensure that the blowout radius satisfies $r_b > |\xi_\mathrm{eq}|$ for any acceleration to be possible \cite{asai2023pathways}. In this new regime, the equilibrium offset $|\xi_\mathrm{eq}| \gtrsim (2c/\omega_{p})\, n_0^{3/2}\, [\SI{e19}{\centi\meter^{-3}}]\, \reviewthree{E^{2}}\, [\SI{10}{\giga e\volt}]\, \reviewthree{\left<x^2\right>_\beta\, [(\SI{5}{\micro\meter})^2]}$, can be comparable to the blowout radius, on the order of $c/\omega_{p}$. This marks the parameter regime where betatron cooling becomes comparable to the acceleration force \cite{lu2007generating}.

\reviewthree{The dynamics of an electron undergoing betatron cooling is described by the scaled amplitude $A$ and phase $\phi$, defined through $A^2=\gamma^{1/2}\left(x^2+2p_x^2/\gamma\right)$ and $x=\gamma^{-1/4}A\cos\phi$. $A^2$ is proportional to the betatron action \cite{goldstein1990classical} and is therefore conserved under acceleration alone; the physical betatron amplitude is $A_\perp=\gamma^{-1/4}A$ (See Supplemental Material for furhter details \cite{SM}). Combining Eqs.~\eqref{eq:dr}--\eqref{eq:dp_perp} gives}
\begin{equation}
    \reviewthree{\frac{dA^2}{dt}=\frac{\dot\gamma}{2\gamma}\,A^2\cos2\phi-\frac{r_e}{3}\,\gamma^{1/2}A^4\cos^2\phi\,\sin^2\phi,}\label{eq:diff_traj}
\end{equation}
\reviewthree{with $\dot\gamma=E-\frac{r_e}{6}\gamma^{3/2}A^2\cos^2\phi$ and $E=-\xi/2$, up to terms of order $p_x^2/\gamma^2$~\cite{Zeng2021}; the phase advances at the betatron frequency $\omega_\beta=\omega_p/(2\gamma)^{1/2}$.} \reviewthree{$A$} is always positive, while the sign of $xp_x$ indicates radial direction of motion: when $xp_x>0$ is outward motion, and $xp_x<0$, is inward. To determine $A(t)$ we use the Krylov--Bogoliubov averaging method \reviewthree{over the fast phase}, assuming that the cooling acts as a perturbation of the betatron trajectory. \reviewthree{The accelerating field drops out of $\langle dA^2/dt\rangle$, while the radiative loss retains its correlation with the phase, $\langle\cos^2\phi\cos2\phi\rangle=1/4$, and} we find that $A(t)=A_0/\sqrt{1+A_0^2\tau}$ \cite{krylov1950introduction}, where $A_0$ is \reviewthree{its initial value} and \reviewthree{$\tau=(r_e/16)\int_0^t\gamma(t')^{1/2}dt'$}, $\tau$ is a normalized time, such that when $\tau=1$ for large $A_0$, the oscillating amplitude is \reviewthree{of order $\gamma^{-1/4}c/\omega_p$}. Importantly, $\tau$ depends on the evolution of $\gamma(t)$. From here on we consider the case where the accelerating field can increase the particle energy in spite of the radiative losses, applicable to current laboratory experiments, \emph{i.e.}, $\gamma(t)=\gamma_0+Et$\reviewthree{, for which $\tau=(r_e/24E)[(\gamma_0+Et)^{3/2}-\gamma_0^{3/2}]$}. This formalism is explored further for the fully general case in the Supplemental Material \cite{SM}.

This perturbed trajectory holds when the betatron frequency is much faster than the cooling rate, \emph{i.e.}, $\omega_\beta > \sqrt{r_e \gamma(dx/dt) x/6}$ \cite{krylov1950introduction}. We also examine the strong cooling regime, where this approximation breaks down, in the Supplemental Material \cite{SM}, see also references \cite{abramowitz1965handbook, rosenzweig1997space}).The solution for $A(t)$ remains valid for $A<18^{1/4}\,r_e^{\reviewthree{-1/2}}\gamma^{\reviewthree{-1/2}}$. \reviewthree{This holds for typical parameters ($E\sim 10\, \mathrm{GeV}$; $A_\perp<50\, \mu\mathrm{m}$ at $n_0=10^{19}\, \mathrm{cm}^{-3}$, $<9\, \mu\mathrm{m}$ at $10^{20}\, \mathrm{cm}^{-3}$).} 
As $\tau$ grows, the amplitude approaches $A\to A_{\rm lim}=\tau^{-1/2}$, demonstrating that the whole phase-space volume is constricted into a region bounded by \reviewthree{$\gamma^{1/2}(x^2+2p_x^2/\gamma)<A_{\rm lim}^2$}.  $A_\mathrm{lim}(t)$ is a limit-cycle orbit towards which all physically relevant trajectories are asymptotically attracted \cite{goldstein1990classical}.

\begin{figure}
    \centering
    \includegraphics[width=\linewidth]{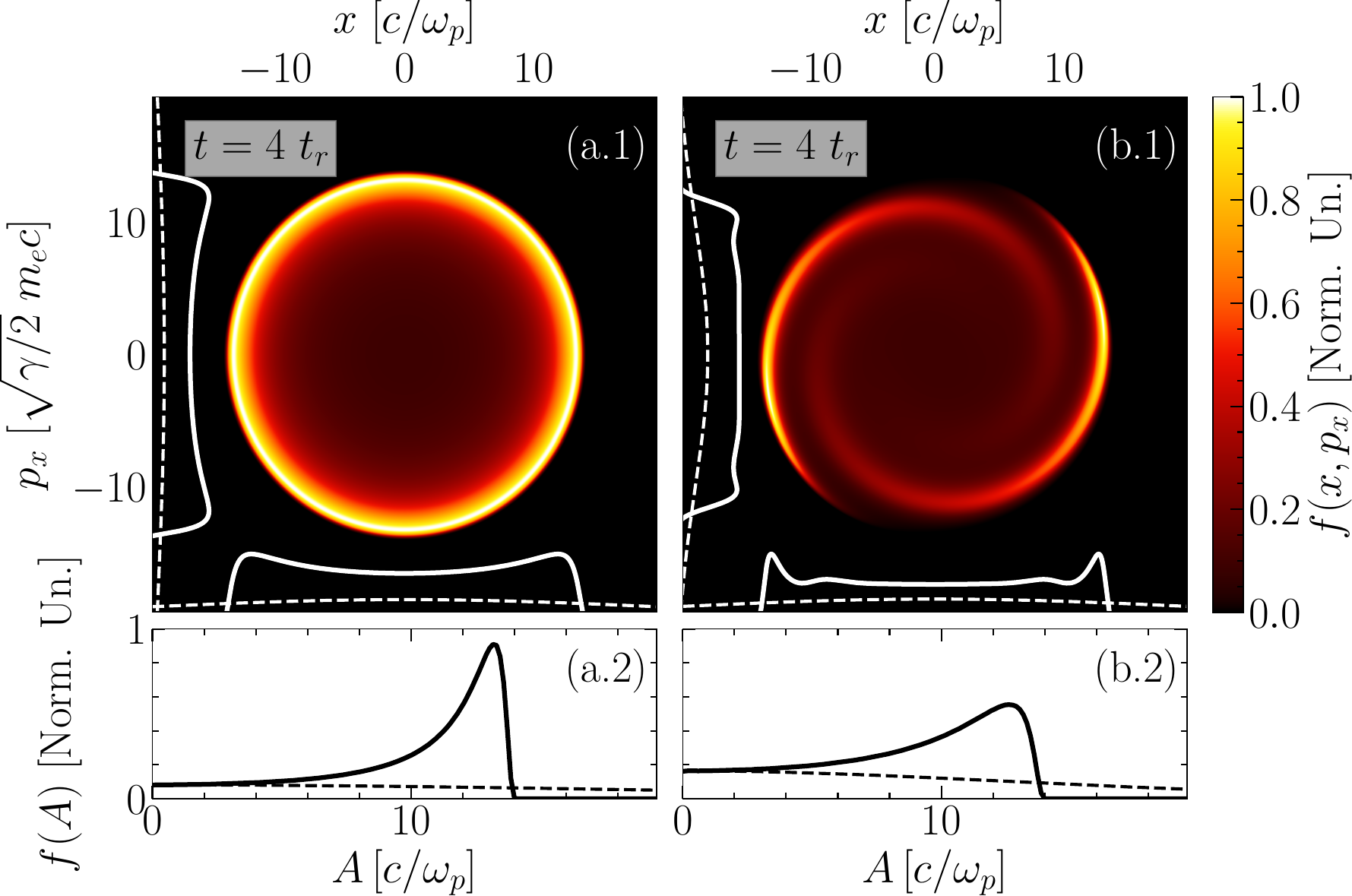}
    \caption{Phase-space evolution governed by Eq.~(\ref{eq:general_sol}), shows how initially smooth Gaussian distributions evolve into ring structures at $t = 4t_r$ (here $\gamma=25$ and $\omega_\beta=10^5\,t_r^{-1}$). Column (a) corresponds to an emittance-matched beam with \reviewthree{$\sigma = 20$ in the normalized units of $A$}; column (b) shows a mismatched beam with \reviewthree{half that momentum spread}. The top row (1) shows the transverse phase-space distribution $f(x, p_x)$, with white lines indicating the projected radial and momentum profiles; dashed lines mark the initial distributions. The second row (2) shows the distribution $f(A)$, highlighting bunching into a narrow peak (solid line). All initial profiles are shown with dashed lines in each panel.} \label{fig:phasespace}
\end{figure}
To describe the kinetic properties of these beams, we introduce the distribution function $f(A,\phi,\gamma)$, with $A$ and $\phi$ related to the transverse displacement and momentum through \reviewthree{$A^2=\gamma^{1/2}(x^2+2p_x^2/\gamma)$} and $\phi=\arctan\!\big((2/\gamma)^{1/2}p_x/x\big)$. From the advection equation in phase-space $df/dt =0$, we obtain the equation for the evolution of $f$ in $A$-$\phi$ phase-space
\begin{equation}
  \frac{\partial f}{\partial t} + \frac{1}{A} \frac{\partial }{\partial A} \left[A \frac{d A(t)}{d t} f\right] + \omega_\beta \frac{\partial f}{\partial \phi} \todiscuss{+E\frac{\partial f}{\partial \gamma}}= 0, \label{eq:distribution_diff}
\end{equation}
and a solution is found with the method of characteristics:
\begin{equation}
  f(A, \phi,t) = \frac{f_0\left(\frac{A}{\sqrt{1-A^2 \tau}}, \phi - t\omega_{\beta}\todiscuss{, \gamma-Et}\right)}{\left(1-A^2 \tau\right)^2}. \label{eq:general_sol}
\end{equation}
The formal similarity between Eq.~\eqref{eq:general_sol} and the synchrotron cooling solution (see Refs.~\cite{bilbao2022radiation, bilbao2024ring}) underscores a key result: betatron-cooled beams also develop population inversions, specifically in the amplitude of their oscillations, \emph{i.e.}, regions where $\partial f /\partial A > 0$. 

Equation \eqref{eq:general_sol} describes the general evolution of any given beam in the blowout structure. We now focus our study on the population inversion process for an initial transverse Gaussian beam profile such that $f_0\propto e^{-A^2/2\sigma^2}$, where $\sigma$ is the beam waist\todiscuss{, and a small spread in energy, with average $\bar\gamma = \gamma_0$, such that we can approximate $f\propto \delta(\gamma_0-\gamma)$}. \todiscuss{By integrating eq.~\eqref{eq:general_sol} in $\gamma$, the resulting $f_A=\int f\, \partial \gamma$} allows to characterize the bunching in phase-space with the formation of a ring of radius $r_r (\tau)$, determined from $\left|\partial_A f_{\todiscuss{A}}\right|_{A=r_r(\tau)} = 0$:
\begin{equation}
    r_r(\tau) = \frac{\sqrt{4\tau \sigma^2-1}}{2\tau \sigma}. \label{eq:ring_rad}
\end{equation}
Equation (\ref{eq:ring_rad}) shows there is a minimum time for the bunching of the amplitudes to appear, $\tau > (4\sigma^2)^{-1}$. The ring radius continues to grow until time $\tau_r=(2\sigma^2)^{-1}$ such that $r_r(\tau_r) \sim \sigma$. This naturally defines the typical timescale for the formation of ring features in the beam distribution as \reviewthree{$t_r=\big[(\gamma_0^{3/2}+12E/(r_e\gamma_0^{1/2}\sigma_\perp^2))^{2/3}-\gamma_0\big]/E$, where $\sigma_\perp=\gamma_0^{-1/4}\sigma$ is the beam waist, which for $Et_r\lesssim\gamma_0$ simplifies to $t_r=8(r_e\gamma_0\sigma_\perp^2)^{-1}$}, or the equivalent propagation length of $L_r\,[\reviewthree{\SI{4.6}{\centi\metre}}]\simeq n_0^{-2}\,[\SI{e19}{\centi\metre^{-3}}]\,E^{-1}\,[\SI{10}{\giga e\volt}]\,\sigma_\perp^{-2}\,[\SI{5}{\micro\metre}]$, realizable in existing laboratory facilities \cite{yakimenko2019facet,joshi2018plasma}. \todiscuss{This timescale can be compared with the driver depletion length $L_d$ \cite{lu2006nonlinear,lu2007generating,SM}. Ring formation requires $L_d \gtrsim L_r$. Using standard blowout-regime estimates, the ratio scales as \reviewthree{$L_d/L_r \propto E^2\,\sigma_\perp^{2}\, n_0^{2}\, n_b^{-1/2}$} \reviewthree{and is $\simeq8$ for the parameters considered here} (see Supplemental Material \cite{SM}). This suggests that the onset of this effect becomes accessible in the emerging regime of multi-GeV beams propagating in high-density plasmas.}

\begin{figure*}[t]
    \centering
    \includegraphics[width = 1\linewidth]{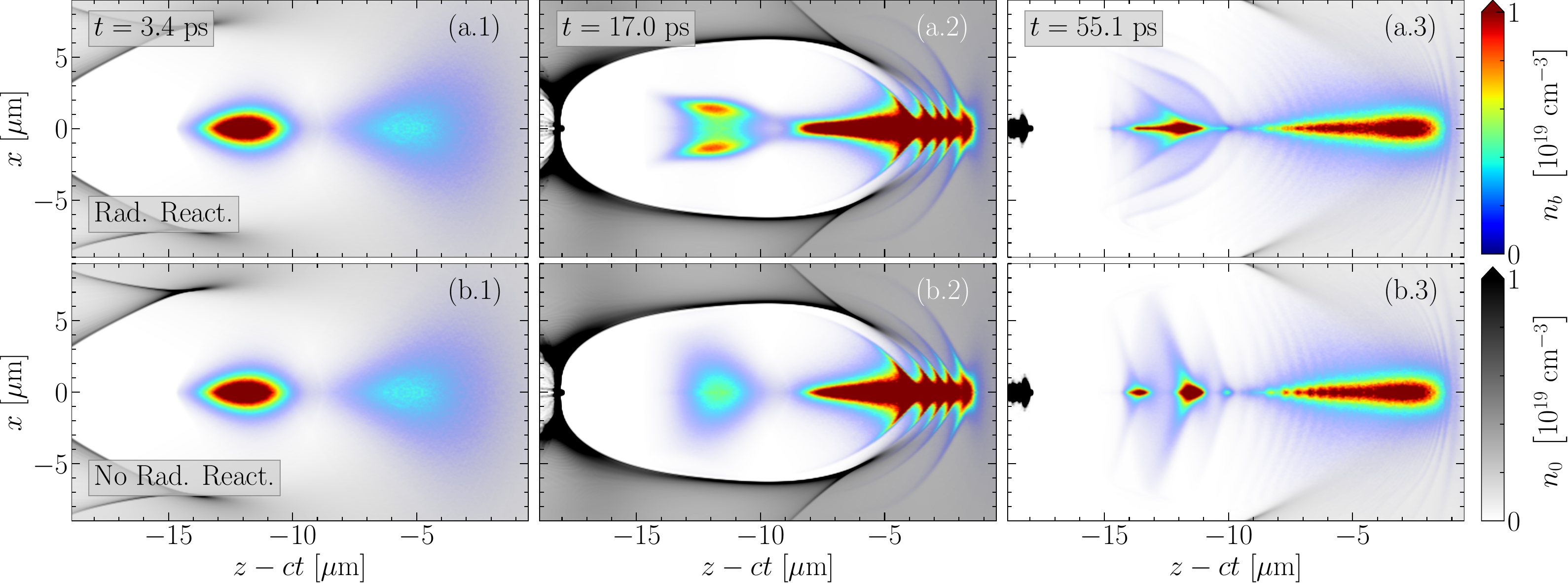}
    \caption{FACET-II–like electron beams develop a transversely bunched profile when propagating through high-density plasma ($n_0 = \SI{1.6e19}{\centi\meter^{-3}}$): a) under the action of radiation reaction, and b) while no bunching appears without radiation reaction. A dense driver beam ($\SI{1}{\nano\coulomb}$, $\sigma_x = \SI{5}{\micro\meter}$, $\sigma_z=\SI{2}{\micro\meter}$) creates a blowout cavity, followed by a trailing beam (\reviewthree{$\SI{0.25}{\nano\coulomb}$}, $\sigma_z=\SI{1}{\micro\meter}$, $\sigma_x=\SI{5}{\micro\meter}$) that undergoes wakefield acceleration. Both beams have a normalized emittance $\varepsilon_n = \SI{300}{\milli\meter\milli\radian}$ and an initial energy of $\SI{10}{\giga\electronvolt}$. The plasma density has a $\SI{2}{\milli\meter}$ upramp (1), $\SI{13}{\milli\meter}$ flat-top (2), and $\SI{2}{\milli\meter}$ downramp (3).}
    \label{fig:ring_FACET2}
\end{figure*}

This amplitude bunching in phase space is illustrated in Fig.~\ref{fig:phasespace} for both emittance-matched and mismatched beams \cite{commentmatching}. In the matched case, radial structuring is clearly visible, appearing as a sharp ring in the transverse phase-space (Fig.~\ref{fig:phasespace}(a.1)) and as a narrow peak in the amplitude distribution $f(A)$ (Fig.~\ref{fig:phasespace}(a.2)). The mismatched beam also shows phase-space bunching signatures, accompanied by phase mixing \cite{clayton2002transverse,joshi2018plasma}. Here, the fine-scale phase-mixed structure in phase space is compressed, and amplitude bunching develops similarly to the matched case, yielding $\partial f/\partial A > 0$ (Fig.~\ref{fig:phasespace}(b.2)). After the development of bunching in amplitude, the spiral structure is still discernible in Fig.~\ref{fig:phasespace}(b.1). Figure~\ref{fig:phasespace}(b.1) (white dashed lines) shows that this single spiral structure translates into multiple peaks in the spatial distribution $f_s(x)$, and multiple pitch-angle bunches in momentum $f_p(p_x)$, where $f_s(x)=\int f(x,p_x)\,dp_x$ and $f_p(p_x)=\int f(x,p_x)\,dx$.

This process is further studied with first principles multi-dimensional \reviewtwo{2D cylindrically symmetric geometry} PIC simulations based on \todiscuss{realistic} beam parameters \reviewtwo{for} FACET-II \todiscuss{beam} \cite{yakimenko2019facet, joshi2018plasma}. Details of the simulation parameters are shown in \reviewthree{Appendix~A}. Figure \ref{fig:ring_FACET2} follows the driver and a witness beam. The driver beam produces a blowout in the upramp region of the background plasma density profile Fig.~\ref{fig:ring_FACET2}(a.1). In this region, the driver transverse envelope decreases due to the focusing forces it self-consistently creates, and the beam density increases, allowing it to drive the blowout structure in the flat-top region. The trailing beam, moving inside the blowout, rapidly develops a ring-shaped transverse structure, as shown in Fig.~\ref{fig:ring_FACET2}(a.2). \todiscuss{This interpretation is further supported by the reconstructed $(x,p_x)$ phase-space from the PIC data, which exhibits the expected ring-like bunching prior to plasma exit, in close agreement with the theoretical predictions for mismatched beams (Fig.~\ref{fig:phase_PIC}, Appendix C). The persistence of this structure after plasma exit, and its manifestation in experimentally accessible screen and spectrometer diagnostics, are also examined in Appendix C.} If the driver beam is long enough, such that the front of the beam drives the wakefield while its rear is experiencing betatron cooling, then a simpler configuration with a single beam that will lead to phase-space bunching is possible, this is shown in the Supplemental Material \cite{SM}.

We also simulate the exit with a down-ramp region: several bunches are observed in the trailing beam, Fig.~\ref{fig:ring_FACET2}(a.3), as modulations of the radial beam profile. This agrees with the predictions of Eq.~\eqref{eq:general_sol}, \emph{cf.} Fig.~\ref{fig:phasespace}(b.1) where non-emittance matched beams show multiple radial spatial bunches due to betatron cooling. For comparison, without radiation reaction, Fig.~\ref{fig:ring_FACET2}(b) shows no signs of bunching or ring formation, confirming that the witness beam profile results from betatron radiation losses.


\reviewtwo{Ring-like beams with $\partial f/\partial A>0$ are a prerequisite for the ICL \cite{whittum1990ion,davoine2018ion}. The beams must also satisfy: i) small relative energy spread, $\Delta\gamma/\gamma < 2\rho/3$, and ii) a narrow distribution of the wiggler parameter $\Delta K/K < \frac{2+K^2}{2K^2}\rho$, where \reviewthree{$K\simeq\langle A\rangle\gamma^{1/4}/\sqrt2$} is the wiggler parameter and $\rho$ is the Pierce parameter \cite{davoine2018ion} (see Supplemental Material for details \cite{SM}). We now show that FACET-II-like beams can become unstable to the ICL. As full-scale self-consistent simulations are computationally prohibitive, we analytically evolve a $\SI{2.5}{\micro\meter}$ witness slice through a wakefield section and a density downramp (Eqs.~\eqref{eq:dr}--\eqref{eq:dp_perp}), and load the resulting phase space into a boosted-frame OSIRIS simulation of the high-density ion channel \cite{davoine2018ion} to fully resolve the self-consistent radiation.}

\begin{figure}
    \centering
    \includegraphics[width=\linewidth]{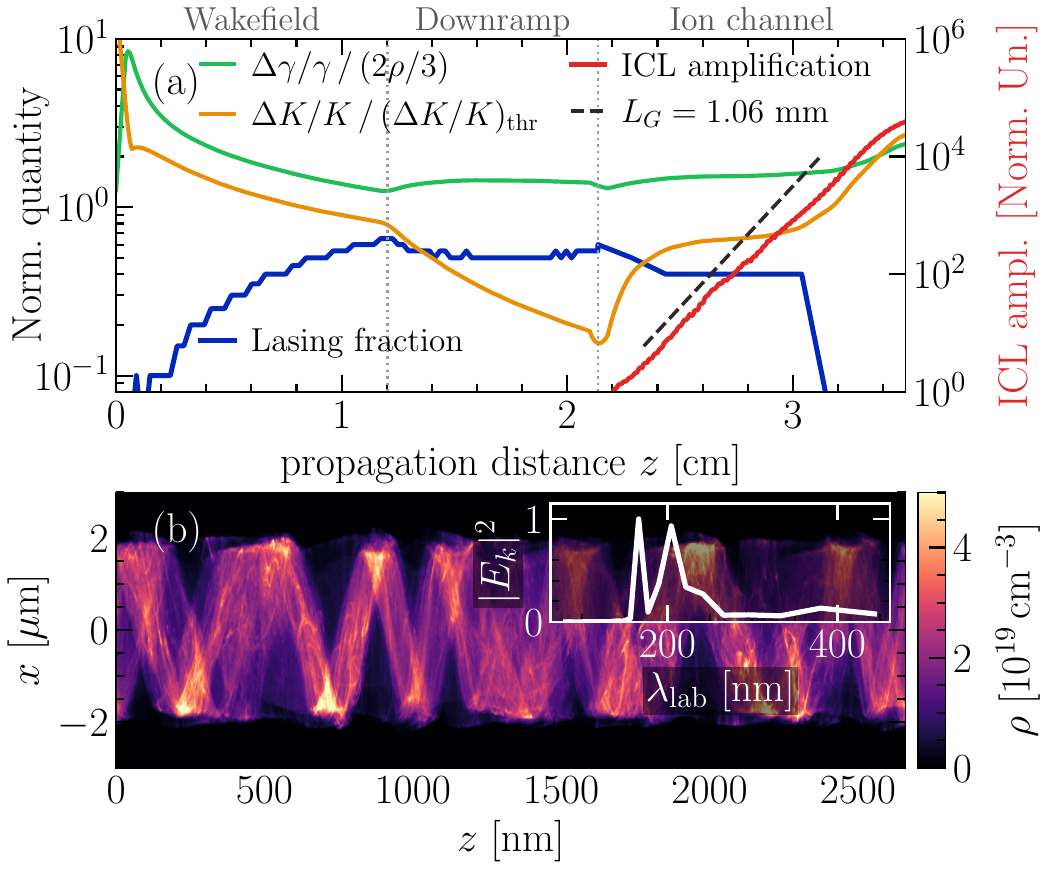}
    \caption{\reviewtwo{ICL onset for a FACET-II-like beam ($E_0=\SI{10}{\giga\electronvolt}$, $\sigma_x=\sigma_z=\SI{5}{\micro\meter}$, $\varepsilon_n=\SI{300}{\milli\meter\milli\radian}$). (a) Energy spread $\Delta\gamma/\gamma/(2\rho/3)$ (green), betatron-strength spread $\Delta K/K/(\Delta K/K)_{\rm thr}$ (orange), and fraction of particles satisfying both lasing conditions (blue) through the wakefield, downramp, and ion channel; the analytical model is followed by a boosted-frame PIC simulation of the radiation growth. Red: radiated field energy; dashed line: 2D gain length $L_G\simeq 1.06$~mm. (b) Charge density at saturation showing microbunching. Inset: spectrum of the transverse $E_2$ field, peaking near $\lambda_{\rm lab}\simeq166$ nm.}}
    \label{fig:ICL}
\end{figure}

\reviewtwo{Figure~\ref{fig:ICL} shows the lasing fraction of the beam reaches $\sim50\%$. The beam microbunches and exponentially amplifies radiation at the resonant wavelength $\lambda_{\rm lab} \simeq \frac{\sqrt{2}\pi c}{\omega_p \gamma^{3/2}}\left(1 + \frac{K^2}{2}\right)\simeq166$ nm. (red curve) \cite{davoine2018ion}. The gain aligns with the theoretical 2D gain length $L_G=2c/(\sqrt{3}\rho_{\rm 2D}\omega_\beta)\simeq \SI{1.06}{\milli\meter}$ \cite{davoine2018ion}. Systematic optimization of this ICL stage will be addressed elsewhere.}

\reviewtwo{This regime is also accessible via LWFA. High-charge ($\sim 1$~nC), $10$-GeV-class LWFA beams \cite{picksley2024matched} could serve as compact drivers to facilitate blowout in subsequent high-density plasmas. Alternatively, petawatt-class lasers \cite{silva2009,sung20100,danson2015petawatt, bromage2021mtw,yoon2021realization} can directly bore ion-channels in dense targets ($n>10^{21}\,\mathrm{cm}^{-3}$) \cite{sarri2012dynamics,schumaker2013ultrafast,russell2025magnetic}. Both scenarios induce extreme radiative losses ($\chi\sim\mathcal{O}(1)$), offering new opportunities to tailor beam injection and phase-space properties \cite{leemans2006gev,kneip2009near,froula2009measurements,vieira2011magnetic,popp2010all, martins2010exploring}.}

\reviewtwo{In conclusion, betatron cooling induces a population inversion in oscillation amplitude, driving phase-space bunching, a fundamentally new kinetic phenomenon in beam-plasma systems representing a laboratory analog to astrophysical synchrotron cooling \cite{bilbao2022radiation,bilbao2024ring, bilbao2024radiative}. These beams are unstable to the ICL, providing an accessible route to coherent radiation generation in ion channels \cite{whittum1990ion,chen1991unified,davoine2018ion}.}

\begin{acknowledgments}
We would like to acknowledge enlightening conversations with Prof. C. Joshi, Prof. W. Mori, Prof. J. Vieira, Dr. B. Malaca,  Prof. R. A. Fonseca, Prof. F. Fiuza,  Dr. P. S. Claveria \& Prof. V. Tikhonchuk. \reviewthree{We thank the anonymous referees, whose thorough and constructive comments led to the inclusion of the ion-channel laser demonstration and significantly improved the analytical description.} This work was supported by FCT (Portugal) (Grant UI/BD/151559/2021, X-MASER - 2022.02230.PTDC \& grant IPFN-CEEC-INST-LA3/IST-ID), and by EuroCC2: funded by the European High-Performance Computing Joint Undertaking (JU) under grant agreement No 101101903.
Simulations were performed at Deucalion (Portugal) funded by FCT Masers in Astrophysical Plasmas 2 (MAPs 2) I.P project 2024.11062.CPCA.A3 (\href{https://doi.org/10.54499/2023.11062.CPCA.A3}{doi:10.54499/2023.11062.CPCA.A3}); and at MareNostrum 5 through the Project 2024.07895.CPCA.A3 (\href{https://doi.org/10.54499/2024.07895.CPCA.A3}{doi:10.54499/2024.07895.CPCA.A3}). 
\end{acknowledgments}
\appendix
\section{End Matter}
\textit{Appendix A: PIC simulation parameters} -- \reviewtwo{Simulations used the OSIRIS particle-in-cell code \cite{fonseca2002osiris} in 2D cylindrical geometry (azimuthal mode $m=0$ \cite{davidson2015implementation}). Data are available on Zenodo (Ref.~\cite{bilbao2025ion}). A longitudinal moving window domain spanned $L_x\times L_r=52\times102\,(c/\omega_p)^2$ with $2500\times5000$ cells ($\Delta x=0.0208\,c/\omega_p$). Radial boundaries were sized to prevent blowout interference. Peak plasma density was $n_0=1.66\times10^{19}\,\mathrm{cm}^{-3}$ ($\omega_p\simeq2.25\times10^{14}\,\mathrm{Hz}$). The time step $\Delta t=0.012\,\omega_p^{-1}$ satisfied the Courant limit.}

\reviewtwo{Gaussian beams were used: a driver ($n_b=0.44\,n_0$, $\SI{1}{\nano\coulomb}$, $\sigma_r=\SI{5}{\micro\meter}$, $\sigma_z=\SI{2}{\micro\meter}$) and a trailing beam ($n_w=0.22\,n_0$, $\sim\SI{0.25}{\nano\coulomb}$, \reviewthree{$\sigma_z\sim\SI{1}{\micro\meter}$}, same $\sigma_r$). Both were initialized with 3D momenta, $10\,\mathrm{GeV}$ energy, $1\%$ energy spread, thermal spread $\Delta p_\perp\sim\reviewthree{60}\,m_ec$, and normalized emittance $\varepsilon_n=\SI{300}{\milli\meter\milli\radian}$.}

\reviewtwo{The background plasma (cold electrons, immobile ions) had a $\SI{2}{\milli\meter}$ linear up-ramp, $\SI{13}{\milli\meter}$ flattop, and $\SI{2}{\milli\meter}$ down-ramp. We used 64 particles/cell, cubic interpolation, and a second-order FDTD solver with standard filtering. Radiation reaction followed the reduced Landau–Lifshitz model \cite{vranic2015particle, vranic2016classical_m, vranic2016quantum_m}.}

\reviewtwo{The lasing beam was analytically tracked [\reviewthree{Eqs.~(2a)--(2c)}] through a wakefield stage ($n_0=\SI{9e19}{\per\centi\meter\cubed}$, $L_\mathrm{WF}=\SI{1.23}{\centi\meter}$), a $\SI{1}{\centi\meter}$ down-ramp, and an ion channel ($n_\mathrm{lase}=\SI{5.0e19}{\per\centi\meter\cubed}$, $L_\mathrm{ch}=\SI{1.4}{\centi\meter}$). Amplification was simulated in a 2D Cartesian slab using a Lorentz-boosted frame ($\gamma_b=31.2$) resolving the $\lambda_1=\SI{161}{\nano\meter}$ line \cite{davoine2018ion}. The cooled beam was initialized as a $16\,\lambda_1$ slice in a static uniform ion column ($n_\mathrm{lase}=\SI{5.0e19}{\per\centi\meter\cubed}$). The domain was $1024\times320$ cells ($64$ cells/$\lambda_1$). The beam ($\sim3\times10^{7}$ macroparticles, quadratic interpolation, peak density $0.44\,n_\mathrm{lase}$) carried $\SI{1}{\nano\coulomb}$ ($\sigma_z=\SI{5}{\micro\meter}$, $\varepsilon_n=\SI{300}{\milli\meter\milli\radian}$). Its lasing core had $\gamma\simeq2.2\times10^{3}$, $K\simeq101$, $\Delta\gamma/\gamma\simeq0.3\%$, and $\Delta K/K\simeq0.3\%$.}

\textit{Appendix B: Parameter scan} -- We perform parameter scans using an idealized simulation setup in which a non-depleting driver beam with $n_b \gg \reviewthree{n_0}$ generates a blowout large enough to accommodate wide witness beams ($\sigma_\perp \sim 10\,\mu$m), enabling direct comparison with the analytical model. The witness beams are emittance-matched, azimuthally symmetric in phase space, and thin longitudinally ($\sigma_\xi \sim 0.1\,c/\omega_p$), consistent with the model assumptions. All beams are positioned downstream of $\xi_\mathrm{eq}$, \reviewthree{so that, after an initial phase in which the wide beams radiate about half of their energy, the accelerating field dominates and the beams gain energy over the remainder of the propagation.}
\begin{figure}
    \centering
    \includegraphics[width = \linewidth]{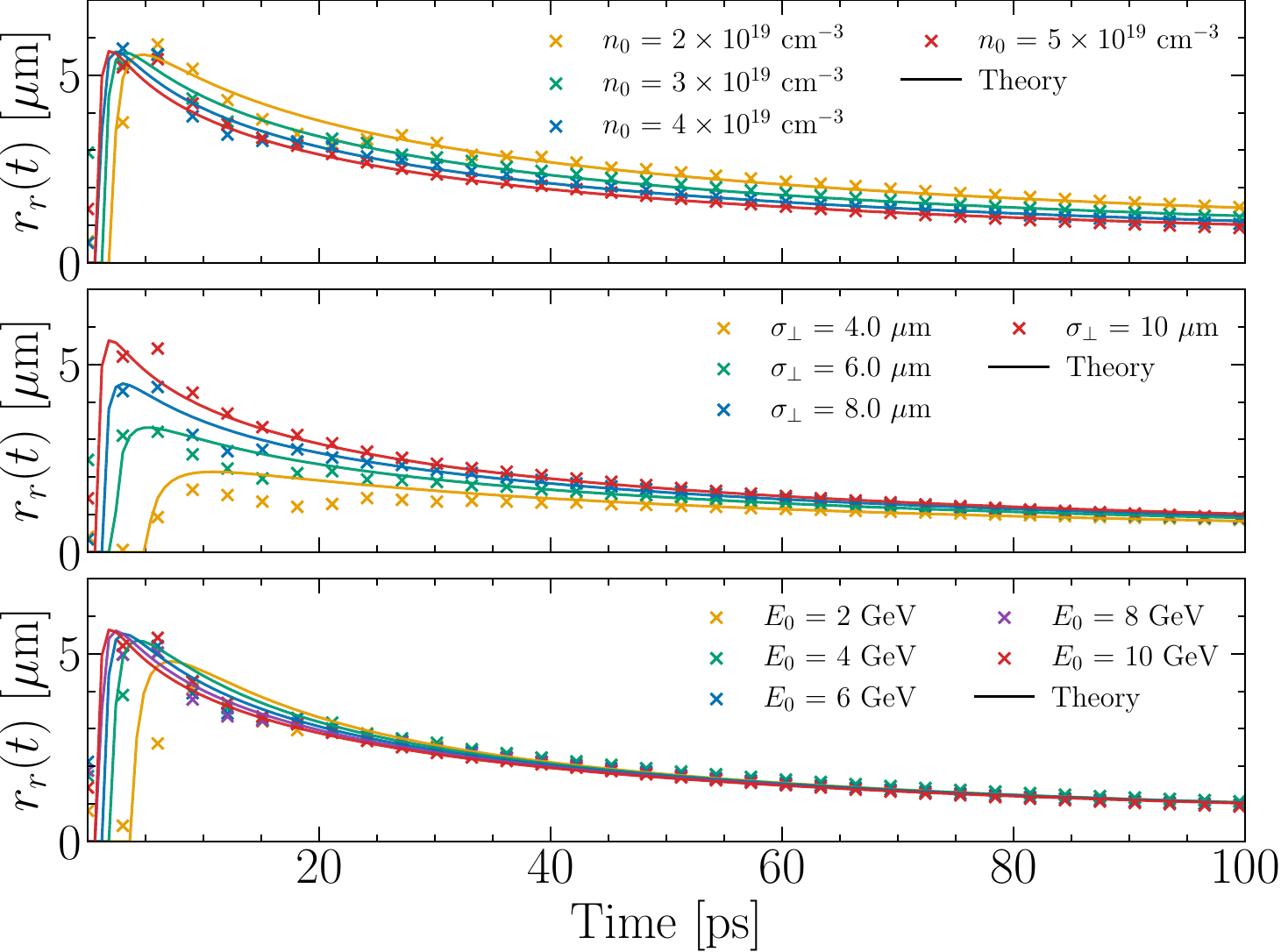}
    \caption{Temporal evolution of the ring radius $r_r(t)$ from particle-in-cell simulations under varied beam and plasma conditions. Rows show scans over plasma density (top), initial transverse beam size (middle), and initial beam energy (bottom), with all other parameters fixed. Markers indicate simulation results and solid lines the analytical prediction [Eq.~(\ref{eq:ring_rad})]\reviewthree{, with $\tau(t)$ evaluated from the measured mean energy $\langle\gamma\rangle(t)$ of the simulated beam and no adjustable parameters}.}
    \label{fig:parameter_scan}
\end{figure}

The results of these parameter scans are shown in Fig.~\ref{fig:parameter_scan}. \reviewthree{The analytical curves are obtained from Eq.~(\ref{eq:ring_rad}) with $\tau(t) = (r_e/16)\int_0^t \langle\gamma\rangle^{1/2}\,dt'$ evaluated from the measured mean energy of the simulated beam, the ring radius in $A_\perp$, $\langle\gamma\rangle^{-1/4} r_r$, and $\sigma_\perp$, $\gamma_0$ and the focusing strength taken from the simulation input; there are no free parameters. The energy history departs strongly from $\gamma_0 + Et$, since the beams first radiate about half of their energy before the accelerating field takes over. In all thirteen runs the model reproduces the time and height of the maximum ring radius and its subsequent decay, with a root-mean-square relative deviation of $5$--$9\%$ for $t > t_r$. Deviations are confined to $t \lesssim t_r$, where the ring is still weakly developed and its radius is not robustly defined.}

\begin{figure}
    \centering
    \includegraphics[width=\linewidth]{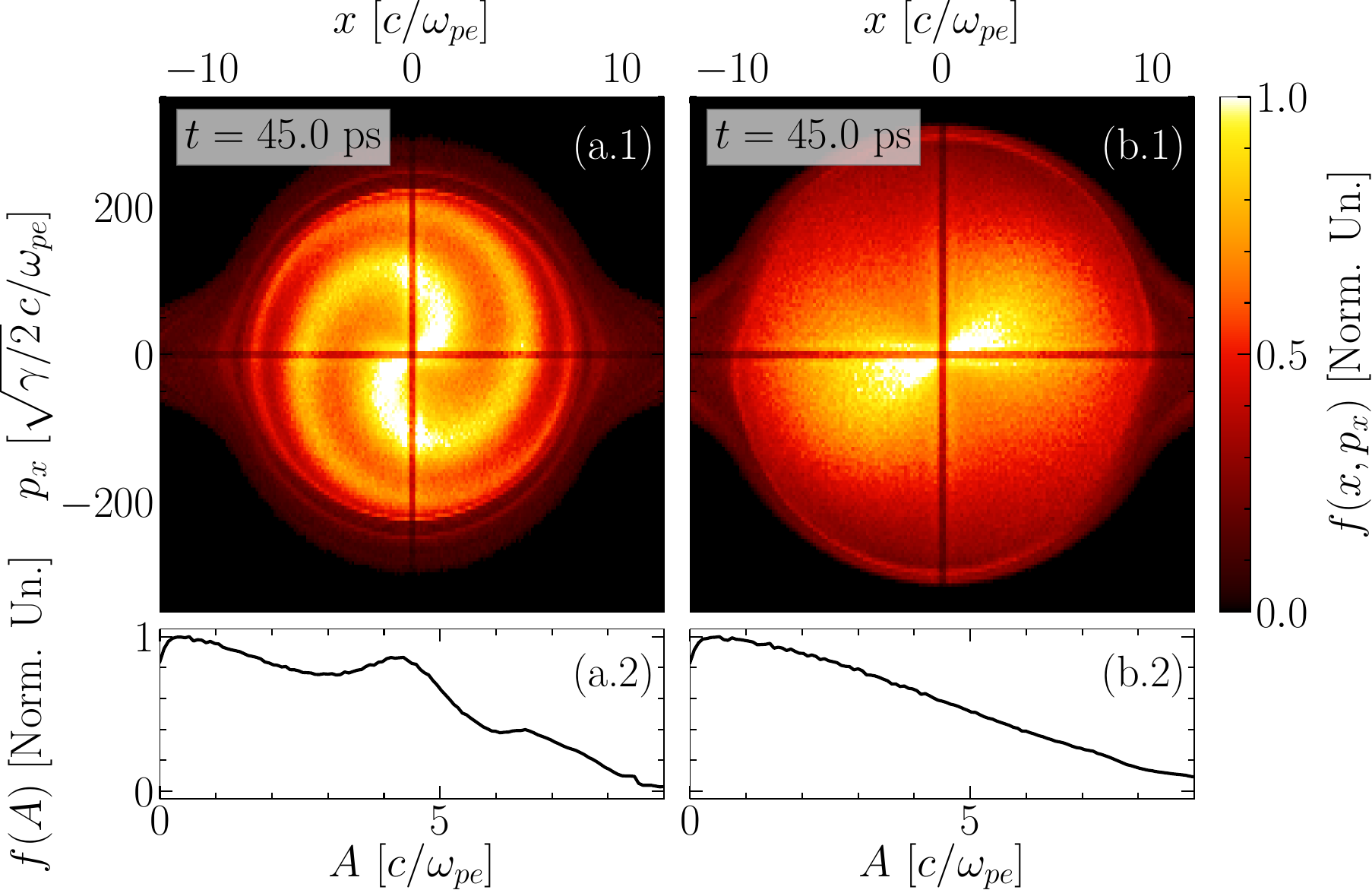}
    \caption{\todiscuss{Reconstructed transverse phase space of the witness beam from PIC data near the end of the high-density plasma region. With radiation reaction (column a), a bunched, ring-like structure appears in $f(x,p_x)$ (row 1), in close agreement with the theoretical prediction for mismatched beams in Fig.~\ref{fig:phasespace}(b.1), and a pronounced peak develops in the distribution $f(A)$ (a.2). Without radiation reaction (column b), neither ring formation nor amplitude bunching is observed.}}
    \label{fig:phase_PIC}
\end{figure}

\begin{figure}
    \centering
    \includegraphics[width=\linewidth]{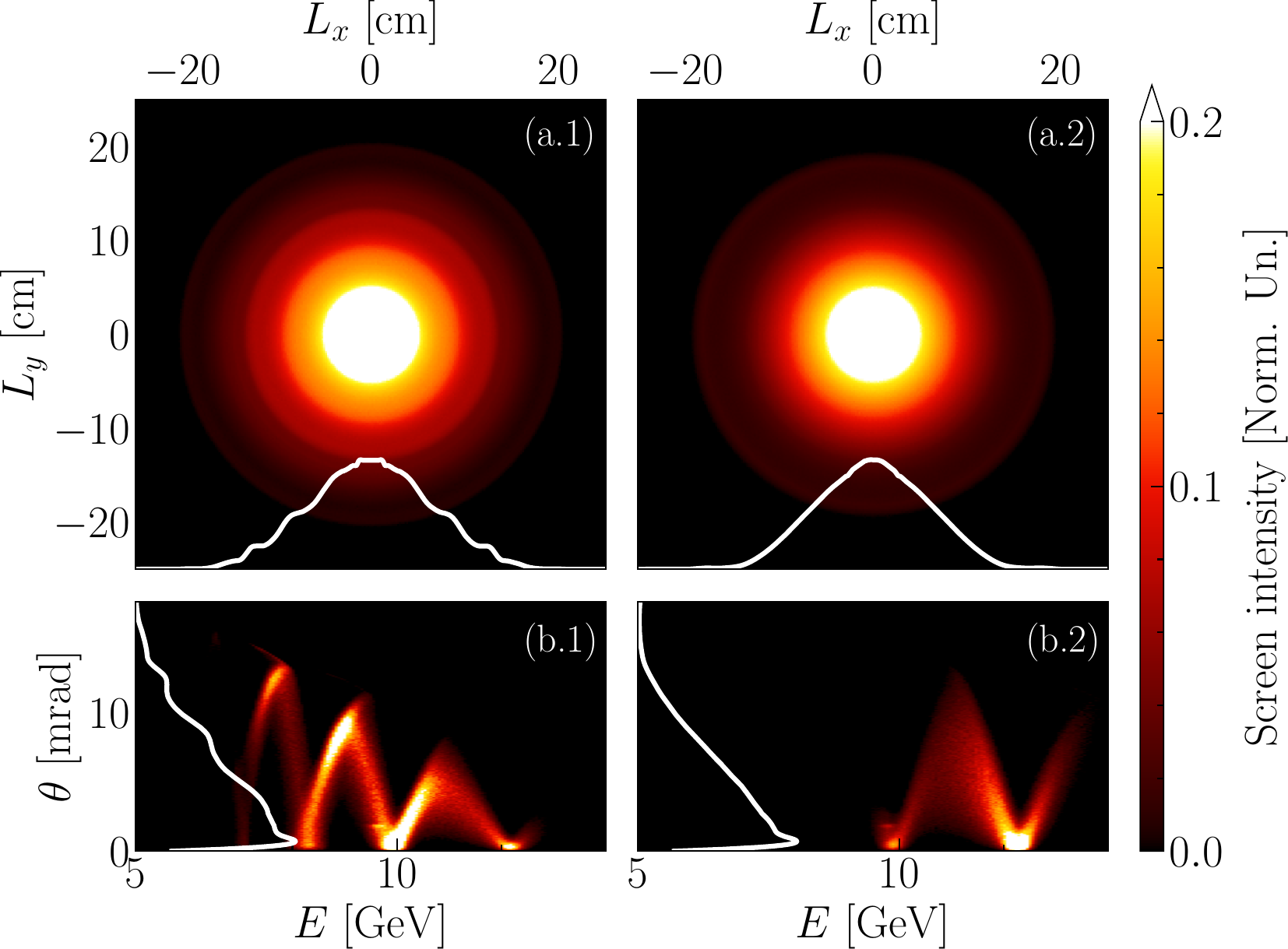}
\caption{\todiscuss{Synthetic luminescent-screen (100 cm downstream) and spectrometer diagnostics after background subtraction, isolating the witness-beam signal. With radiation reaction enabled, clear ring structures appear in the screen image (a.1) and correlate with divergence–energy features (b.1). Without radiation reaction, no ring formation or correlated structure is observed (a.2, b.2).}}
    \label{fig:screen_spec}
\end{figure}

\textit{\todiscuss{Appendix C: Phase-space reconstruction and synthetic diagnostics --}}
\reviewthree{Figure~\ref{fig:phase_PIC} shows the transverse phase space $f(x,p_x)$ and the amplitude distribution $f(A)$ of the witness beam, reconstructed from the PIC data of Fig.~\ref{fig:ring_FACET2} at the end of the flat-top region. With radiation reaction, the beam develops the ring-like bunching and the peak in $f(A)$ predicted for mismatched beams [Fig.~\ref{fig:phasespace}(b)]; without radiation reaction, neither feature appears.}
\todiscuss{Synthetic luminescent-screen and spectrometer diagnostics (100 cm downstream) were computed to assess observability of phase-space bunching. With radiation reaction enabled, ring structures appear in the background-subtracted screen images and correlate with divergence–energy features; without radiation reaction no such structures are observed (Fig.~\ref{fig:screen_spec}). Details of the synthetic diagnostic model are provided in the Zenodo \cite{bilbao2024radiativePIC}.}

\bibliography{apssamp}
\clearpage
\begin{widetext}
\section{Supplementary material}

\section{Simulation results with a single beam}
\begin{figure*}[!h]
    \centering
    \includegraphics[width = 1\linewidth]{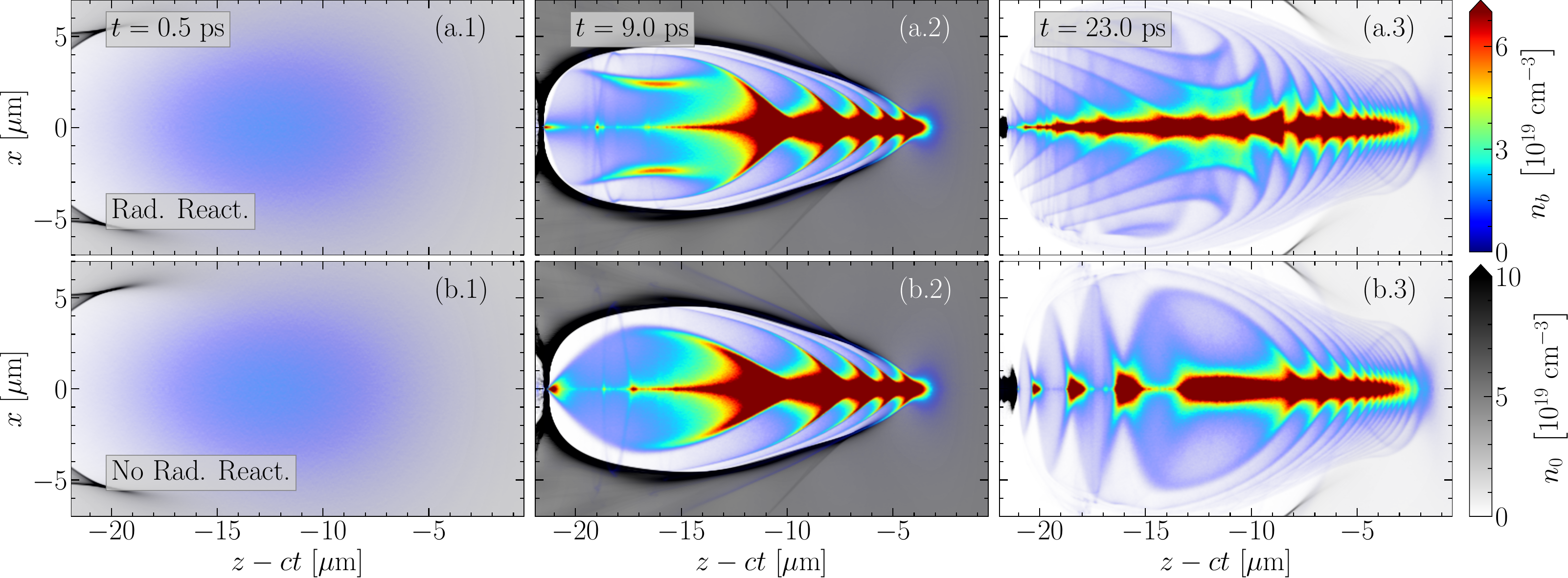}
    \caption{PIC simulations showing \reviewthree{the} formation of ring-shaped, phase-bunched profiles in a single FACET-II–like electron beam propagating through high-density plasma. The front of the beam drives the blowout cavity, while the rear undergoes betatron cooling. Row (a) includes radiation reaction; row (b) does not. Columns correspond to different propagation stages: upramp (1), uniform plasma (2), and downramp (3). Ring formation occurs only when radiation reaction is included.}
    \label{fig:ring_FACET2_one}
\end{figure*}
While the main text focused on a two-beam setup, consisting of a driver and a trailing witness beam, we observed that the tail of the driver also exhibited signatures of betatron-induced ring formation. Motivated by this, we explored a simpler configuration where a single beam simultaneously drives the blowout structure and experiences betatron cooling.

We conducted \reviewthree{axisymmetric cylindrical ($r$--$z$, $m=0$)} PIC simulations using a single electron beam with parameters consistent with FACET-II: total charge of \SI{3}{\nano\coulomb}, energy of \SI{10}{\giga\electronvolt}, normalized emittance of $\varepsilon = \SI{300}{\milli\meter\milli\radian}$, waist of $\sigma_\perp = \SI{5}{\micro\meter}$, and a  $\sigma_z = \SI{5}{\micro\meter}$, with peak density of $n_b = 9.5\times10^{18}\, \mathrm{cm}^{-3}$. We employ the same profile as in the case with 2 beams. The beam propagates through a plasma with peak density $5\times10^{19}\, \mathrm{cm}^{-3}$. The rest of the parameters match those used in the two-beam scenario presented in the main body of the manuscript, to allow for direct comparison. The main difference is that here the driver beam is longer and there is no witness beam.

The simulation results, shown in Fig.~(\ref{fig:ring_FACET2_one}), confirm that the beam drives a blowout cavity in the upramp region (a.1), while the rear of the beam develops a clear ring-shaped structure as it propagates through the uniform-density plasma (a.2). These features persist as the beam exits into the downramp region (a.3), confirming the robustness of the phase-space bunching. In contrast, when radiation reaction is disabled (row b), the beam remains smooth and featureless throughout propagation.

\begin{figure}[!h]
    \centering
    \includegraphics[width=0.6\linewidth]{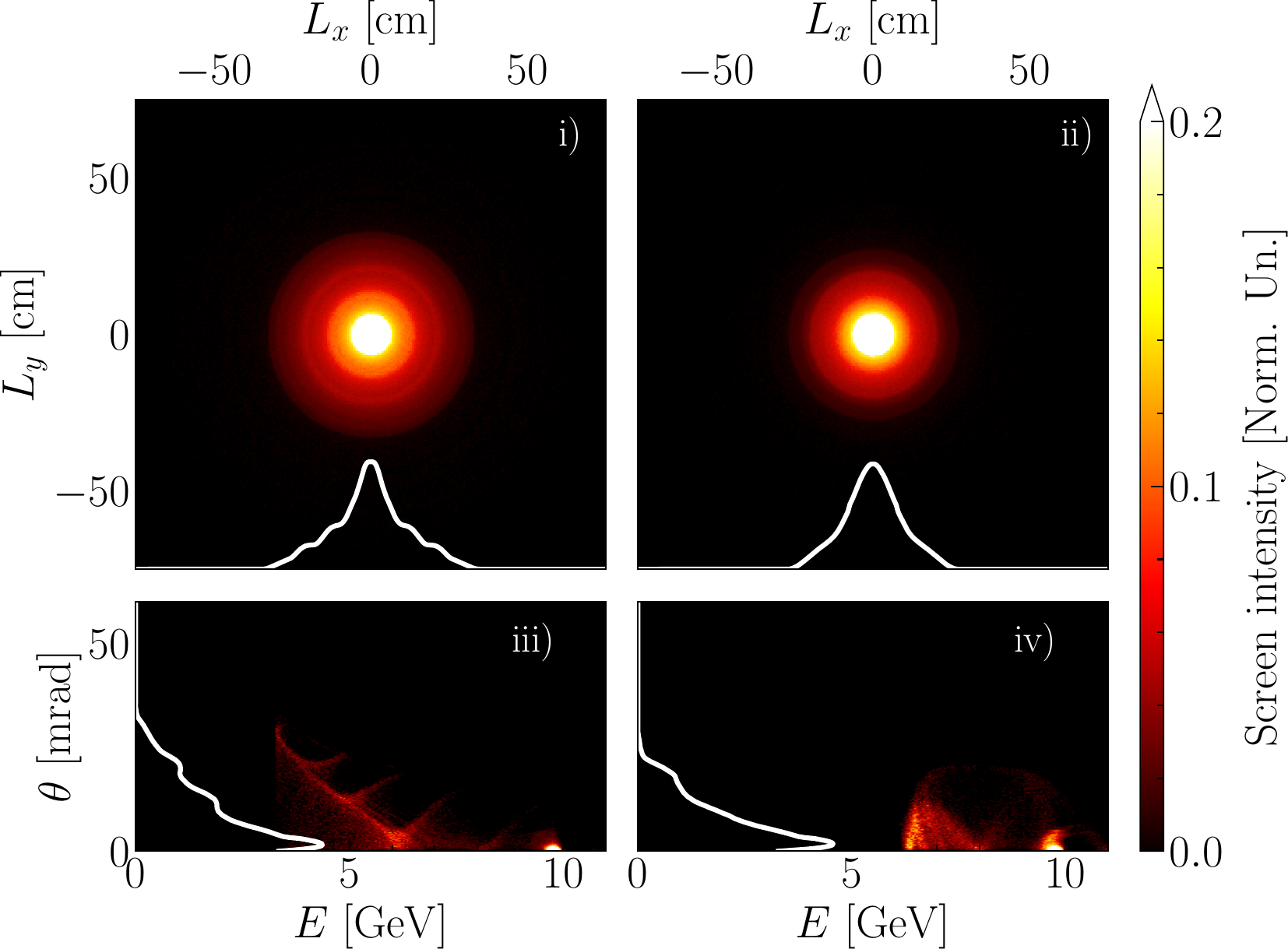}
    \caption{Synthetic diagnostics for the single-beam configuration. (i) Downstream screen (100 cm from plasma exit) with radiation reaction shows visible ring structures. (ii) Without radiation reaction, the screen shows no ring formation. (iii) Divergence versus energy plot with radiation reaction reveals distinct ring-associated features. (iv) Without radiation reaction, the divergence spectrum is smooth and unstructured.}
    \label{fig:screen_spec_one}
\end{figure}
To assess the experimental observability of these structures, we simulated downstream diagnostics analogous to those presented in the main text. A synthetic screen placed \SI{100}{\centi\meter} downstream from the plasma exit (Fig.\ref{fig:screen_spec_one}.i) clearly reveals ring-shaped intensity patterns in the presence of radiation reaction. The corresponding energy-resolved divergence plot (Fig.\ref{fig:screen_spec_one}.iii) shows that each ring possesses distinct energy and emittance characteristics, again confirming the presence of phase-space bunching. In contrast, simulations without radiation reaction produce neither ring structures nor energy-dependent features, see Figs.~(\ref{fig:screen_spec_one}.ii \& \ref{fig:screen_spec_one}.iv).

These results demonstrate that ring-shaped, phase-structured beams can emerge naturally in single-beam setups and are robust against diagnostic and propagation effects. This provides a more accessible experimental configuration, requiring only a single high-energy beam to observe the effects of betatron cooling.

\section{Synthetic screen and spectrometer diagnostics from PIC simulations}
We export the full macroparticle data at the final simulation iteration, i.e., positions $\mathbf{r}_i$, momenta $\mathbf{p}_i$, and weights $w_i$. After the beam has transitioned to vacuum, particles are propagated ballistically to a detection plane placed $100$ cm downstream. For each particle, the intersection time is $t_i^\ast=(z_\mathrm{scr}-z_i)/v_{z,i}$, after which $x_i^\ast=x_i+v_{x,i} t_i^\ast$ and $y_i^\ast=y_i+v_{y,i} t_i^\ast$ are recorded (constant momentum throughout). All particles that reach the plane are included. The screen image is formed by depositing the particle weights $w_i$ onto a regular $(x,y)$ grid at $(x_i^\ast,y_i^\ast)$; intensities are normalized to the peak, and we display the central window $x,y\in[-25,25]$ (cm in the figures).

For the synthetic spectrometer, we compute for each particle the divergence $\theta_i=\tan^{-1}(p_{\perp,i}/p_{\parallel,i})$ and energy $E_i=\gamma_i m_e c^2$, then accumulate $w_i$ on a regular grid in $(\theta,E)$. No instrumental response, filtering, or resolution broadening is applied; the maps reflect direct projections of the exported simulation data. To isolate the witness signal, particles tagged as driver are excluded prior to both projections. Experimentally, this corresponds to subtracting driver-only background shots from shots containing both driver and witness. For the case of a single bunch, no subtraction is necessary.

\newpage
\section{General averaged dynamics and cooling regimes}
In this section we present the general averaged description of betatron cooling, identify the invariant structure of the dynamics, and organize the physically relevant regimes. This provides a unified framework for the simplified solutions employed in the main text and clarifies the relation between ion-channel cooling, slow cooling under acceleration, and the late-time asymptotic attractor.

\subsection{Averaged equations of motion}
Starting from the normalized equations of motion including radiation reaction (see main text),
\begin{align}
\dot x &= \frac{p_x}{\gamma}, \label{eq:x}\\
\dot p_x &= -\frac{1}{2}x - \frac{r_e}{6}\,\gamma p_x x^2, \label{eq:px}\\
\dot\gamma &= E - \frac{r_e}{\reviewthree{6}}\,\gamma^2 x^2 , \label{eq:gamma}
\end{align}
we assume rapid betatron oscillations and $p_x/\gamma\ll1$. \reviewthree{As in the main text, we describe the orbit by the scaled amplitude $A$ and phase $\phi$,
\begin{equation}
A^2 = \gamma^{1/2}\Big(x^2+\frac{2p_x^2}{\gamma}\Big),\qquad
x=\gamma^{-1/4}A\cos\phi,\qquad
p_x=-\frac{\gamma^{1/4}}{\sqrt2}\,A\sin\phi ,
\label{eq:actionangle}
\end{equation}
where $A$ is measured in units of $\gamma^{1/4}c/\omega_p$. The pair $(\phi,A)$ is not canonical; the action conjugate to $\phi$ is $J=\frac{1}{2\pi}\oint p_x\,dx=A^2/(2\sqrt2)$ (in units of $m_ec^2/\omega_p$), so $A^2$ is proportional to the betatron action and is an adiabatic invariant under acceleration. Differentiating $A^2$ along the exact motion, Eqs.~\eqref{eq:x}--\eqref{eq:gamma}, the Lorentz-force terms cancel identically (they only advance $\phi$), leaving
\begin{equation}
\frac{dA^2}{dt}=\frac{\dot\gamma}{2\gamma}\,A^2\cos2\phi-\frac{r_e}{3}\,\gamma^{1/2}A^4\cos^2\phi\sin^2\phi,
\qquad
\dot\gamma = E-\frac{r_e}{6}\,\gamma^{3/2}A^2\cos^2\phi ,
\label{eq:exactA}
\end{equation}
up to terms of order $p_x^2/\gamma^2$, which average to zero.} Applying Krylov--Bogoliubov averaging over one betatron period \reviewthree{at fixed $A$, the accelerating field drops out of $\dot A$, since acceleration conserves the action, while the radiative part of $\dot\gamma$ retains its correlation with the phase, $\langle\cos^2\phi\cos2\phi\rangle=1/4$. This} yields the reduced system
\begin{equation}
\reviewthree{\dot A = -\frac{r_e}{32}\,\gamma^{1/2}A^3,
\qquad
\dot\gamma = E - \frac{r_e}{12}\,\gamma^{3/2} A^2 .}
\label{eq:avg}
\end{equation}
\reviewthree{In terms of the betatron amplitude $A_\perp=\gamma^{-1/4}A$, Eq.~\eqref{eq:avg} reads $\dot A_\perp=-(E/4\gamma)A_\perp-(r_e/96)\gamma A_\perp^3$: the first term is the adiabatic damping under acceleration and the second the nonlinear radiative damping responsible for phase-space bunching. Neglecting the contribution of $\dot\gamma$ to $\ddot x$ omits the $\cos2\phi$ term in Eq.~\eqref{eq:exactA} and overestimates the radiative damping of $A_\perp$ by a factor of two. We have verified Eq.~\eqref{eq:avg} against direct numerical integration of Eqs.~\eqref{eq:x}--\eqref{eq:gamma} for $E=0$ and for $E$ equal to one and five times the initial radiative loss rate, finding agreement to better than $10^{-3}$ over 120 betatron periods.} These equations govern the slow evolution of amplitude and energy in all regimes discussed below.

\subsection{Invariant and general (implicit) solution}
Eliminating time between the averaged equations \eqref{eq:avg} gives a first integral relating $\gamma$ and $A$,
\begin{equation}
\reviewthree{\gamma^{3/2}(A) = C A^4 + \frac{8E}{r_e}\,A^{-2},}
\label{eq:invariant}
\end{equation}
where the constant $C$ is fixed by the initial condition $(A_0,\gamma_0)$, \reviewthree{$C=(\gamma_0^{3/2}-\frac{8E}{r_e}A_0^{-2})/A_0^4$}, and is a constant of motion. Equation \eqref{eq:invariant} parametrizes trajectories in the $(A,\gamma)$ plane, shown in Fig.~(\ref{fig:trajectories_gamma}).
\begin{figure}[!h]
    \centering
    \includegraphics[width=0.85\linewidth]{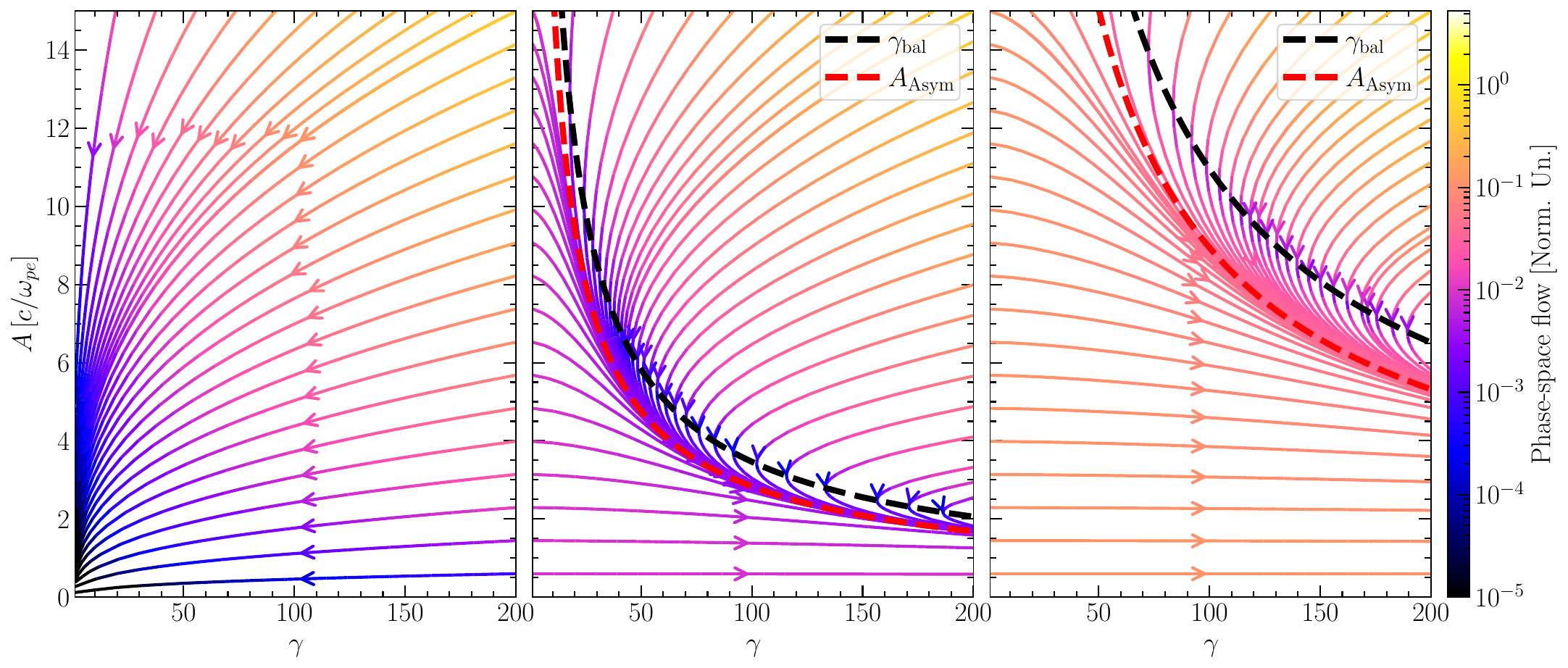}
    \caption{\reviewthree{Phase-space stream plots of the reduced averaged system~\eqref{eq:avg} in the $(\gamma,A)$ plane, with $A$ in units of $\gamma^{1/4}c/\omega_{p}$ and time normalized to $\omega_{p}^{-1}$. Three values of the accelerating field $E$ (in units of $m_ec\,\omega_p$) are shown, while the radiation-reaction strength $r_e=10^{-4}\,c/\omega_{p}$ is held fixed. For $E=0$ (left panel) the energy collapses while $A$ changes only slowly through the cubic damping term, following $\gamma^{3/2}\propto A^4$, as expected for an ion channel. For finite $E=0.1$ and $E=1$ (middle and right panels, respectively), phase space separates into regions of initial radiative cooling and subsequent acceleration. Trajectories that begin above the instantaneous balance curve (black dashed line), $\gamma_{\rm bal}=(12E/r_e)^{2/3}A^{-4/3}$, initially decelerate and lose amplitude along this curve before crossing into the accelerating regime, whereas trajectories below it accelerate immediately at nearly constant $A$. The red dashed line is the self-similar attractor of Eq.~\eqref{eq:asymptotics_box}, the $C=0$ curve $\gamma^{3/2}=(8E/r_e)A^{-2}$ of Eq.~\eqref{eq:invariant}; all accelerating trajectories approach $A\propto t^{-3/4}$, $\gamma\propto t$ at late times, beyond the range of $\gamma$ shown.}}
    \label{fig:trajectories_gamma}
\end{figure}

Substituting Eq.~\eqref{eq:invariant} into the amplitude equation yields the general implicit solution
\begin{equation}
t
= -\frac{\reviewthree{32}}{r_e}
\int_{A_0}^{A(t)}
\frac{dA}{A^3\reviewthree{\left(C A^4 + \dfrac{8E}{r_e}A^{-2}\right)^{1/3}}}.
\label{eq:quadrature}
\end{equation}
This quadrature is non-elementary in general and defines the full solution implicitly; closed-form expressions arise only in specific limits discussed below. \reviewthree{Alternatively, for any prescribed energy history $\gamma(t)$, the amplitude equation in \eqref{eq:avg} integrates directly to
\begin{equation}
A(t)=\frac{A_0}{\sqrt{1+A_0^2\,\tau(t)}},\qquad
\tau(t)=\frac{r_e}{16}\int_0^t\gamma(t')^{1/2}\,dt',
\label{eq:closed}
\end{equation}
which is the solution used in the main text, and the form used for comparison with the simulations, where $\gamma(t)$ is taken from the PIC data.}
\subsection{Exactly solvable case: ion-channel (pure cooling)}
In the absence of acceleration ($E=0$), Eq.~\eqref{eq:invariant} reduces to \reviewthree{$\gamma^{3/2} = C A^4$}, and the averaged equations admit an exact solution,
\begin{equation}
\reviewthree{A(t)=A_0\left(1+\kappa t\right)^{-3/10},
\qquad
\gamma(t)=\gamma_0\left(1+\kappa t\right)^{-4/5},
\qquad
\kappa=\frac{5}{48}\,r_e\gamma_0^{1/2}A_0^2 .}
\label{eq:ionchannel}
\end{equation}
Thus, in the ion-channel limit \reviewthree{$A\propto t^{-3/10}$ and $\gamma\propto t^{-4/5}$}; these trajectories describe pure radiative cooling without energy gain and provide an exact solution. \reviewthree{In terms of the betatron amplitude, $A_\perp=\gamma^{-1/4}A\propto t^{-1/10}$, i.e.\ $\gamma\propto A_\perp^8$: the energy collapses much faster than the spot size, because radiation does not alter the orbit angle and the amplitude shrinks only through the stiffening of the focusing force. The normalized emittance $\varepsilon_n\propto\gamma^{1/2}A_\perp^2=A^2\propto\gamma^{3/4}$ is thus damped mainly by energy loss.}
\subsection{Early-time regimes for $E\neq0$}
When acceleration is present ($E\neq0$), two distinct early-time regimes may occur, depending on the relative strength of radiation reaction and acceleration.
\paragraph{(i) Slow cooling: acceleration-dominated regime.}
If radiative losses are initially weak, \reviewthree{$\frac{r_e}{12}\gamma^{3/2} A^2 \ll E$}, the energy evolution is approximately $\gamma(t)\simeq \gamma_0 + E t$. Substituting this into \reviewthree{Eq.~\eqref{eq:closed}} yields the slow-cooling solution used throughout the main text,
\begin{equation}
\reviewthree{A(t)=\frac{A_0}{\sqrt{1+A_0^2\tau(t)}},
\qquad
\tau(t)=\frac{r_e}{24E}\left[(\gamma_0+Et)^{3/2}-\gamma_0^{3/2}\right]\simeq\frac{r_e}{16}\gamma_0^{1/2}t .}
\end{equation}
This regime corresponds to gradual emittance damping over many betatron oscillations and leads to the phase-space bunching and ring formation analyzed in the main paper.
\paragraph{(ii) Fast cooling: overdamped early evolution.}
For particles injected with sufficiently large initial amplitude, the radiative term initially dominates the longitudinal dynamics, \reviewthree{$\frac{r_e}{12}\gamma^{3/2} A^2 \gg E$}, so that energy gain from the wakefield is negligible over the initial cooling phase. In this regime the averaged equations \eqref{eq:avg} reduce to
\begin{equation}
\reviewthree{\dot A \simeq -\frac{r_e}{32}\,\gamma^{1/2}A^3,
\qquad
\dot\gamma \simeq -\frac{r_e}{12}\,\gamma^{3/2} A^2 ,}
\label{eq:fast_cooling}
\end{equation}
which are identical to the ion-channel (pure cooling) equations up to subleading corrections; consequently the early-time evolution follows the ion-channel trajectories \reviewthree{$A(t)\propto t^{-3/10}$ and $\gamma(t)\propto t^{-4/5}$} until the amplitude is sufficiently reduced that \reviewthree{$\frac{r_e}{12}\gamma^{3/2} A^2 \sim E$}, at which point the dynamics crosses over smoothly into the slow-cooling, acceleration-dominated regime.
\subsection{Late-time asymptotics (global attractor)}
Starting from the averaged equations \eqref{eq:avg}, \reviewthree{we seek a self-similar late-time solution of the form $A=k\,t^{a}$, $\gamma=\kappa Et$. The amplitude equation fixes $a=-3/4$ and $k^2=24/(r_e\sqrt{\kappa E})$, and the energy equation then gives $\kappa=1/3$. This} yields the universal asymptotic laws
\begin{equation}
\reviewthree{A(t)\simeq \left(\frac{24\sqrt3}{r_e\sqrt{E}}\right)^{1/2}t^{-3/4},\qquad
\gamma(t)\simeq \frac{E}{3}\,t ,}
\label{eq:asymptotics_box}
\end{equation}
which are independent of the initial amplitude and describe the global attractor of the averaged dynamics\reviewthree{; it corresponds to $C=0$ in Eq.~\eqref{eq:invariant}, and to $A_\perp\simeq\sqrt{72/(r_eE)}\,t^{-1}$ for the betatron amplitude. On the attractor the particle retains one third of the energy gained from the wake and radiates the remainder, so the late-time state is a balance between gain and loss rather than the acceleration-dominated limit}.

\section{Strongly cooled regime}
There exists a second regime for betatron cooling, where initially the perpendicular radiative force is much larger than the perpendicular Lorentz force. This remains within the range of validity of the Landau--Lifshitz model as long as the total Lorentz force exceeds the radiation reaction force. The transition between the slow and strong cooling regimes is determined by this classical balance between the perpendicular Lorentz and radiative forces, not by the quantum parameter $\chi$. Both regimes can, in principle, occur for either $\chi \ll 1$ (classical) or $\chi \gtrsim 1$ (quantum) depending on beam energy and plasma density. The analysis presented here employs the Landau–-Lifshitz formulation and therefore applies to the classical limit, but the same qualitative behavior, rapid amplitude collapse and phase-space bunching, is expected to persist when quantum, corrected radiation reaction models are used for $\chi \gtrsim 1$.

\begin{figure}[!h]
    \centering
    \includegraphics[width=0.85\linewidth]{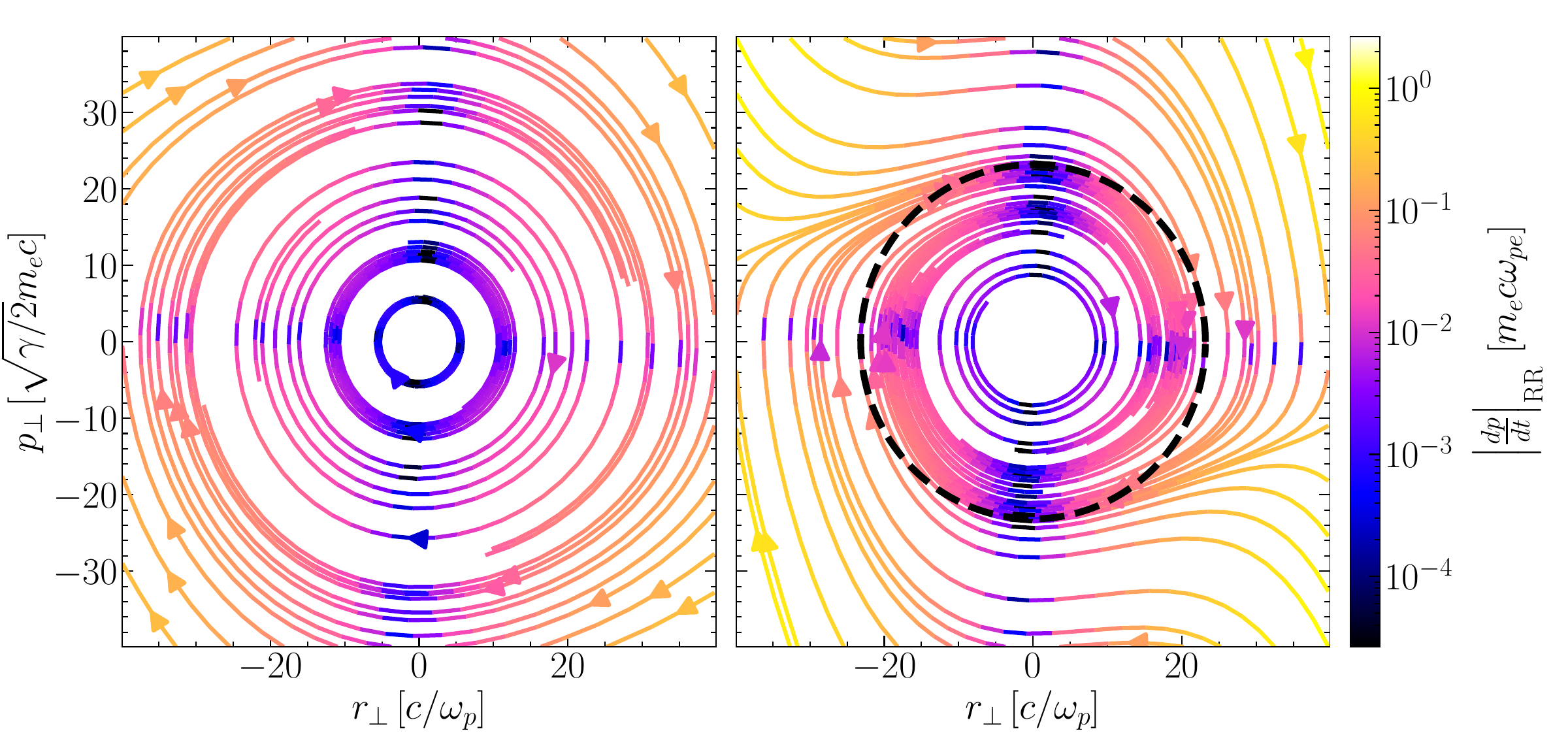}
    \caption{Phase-space stream plots, obtained from \reviewthree{the vector field $(\dot r_\perp,\dot p_\perp)$ of Eqs.~(2a) and (2c) in the main text} (for a constant $p_\xi = 1000\, m_e c$) demonstrate the bunching in phase-space as a result of betatron cooling and showcase the different regions of phase-space where the particles undergo either underdamped oscillations or overdamped oscillations.  The left plot shows the slow cooling regime, where radiative losses are small compared with \reviewthree{the focusing force}, here $r_e^2\omega_p^2/c^2 = 10^{-4}$. The \reviewthree{right} plot shows streamlines where the radiative losses are strong and rapidly collapse the phase-space if the amplitude of oscillation is too large, here $r_e^2\omega_p^2/c^2 = 5\times10^{-4}$. The region enclosed by the black dashed circle corresponds to the phase-space boundary $\reviewthree{A_\perp}<A_b=\sqrt{3\sqrt{2} }/(\reviewthree{r_e^{1/2}} \gamma^{3/4})$, the region where the cooling is slow.}
    \label{fig:trajectories_ion}
\end{figure}

To access this regime the betatron amplitude \reviewthree{$A_\perp$} is larger than $A_b\reviewthree{=\sqrt{3\sqrt{2} }/(r_e^{1/2} \gamma^{3/4})}$, where we recall that amplitude $\reviewthree{A_\perp}$ and the classical electron radius $r_e$ are normalized to the background plasma \reviewthree{skin depth}.  For example, in a plasma with density $\SI{e19}{\centi \metre^{-3}}$, a $\SI{10}{\giga e\volt}$ beam must have a transverse size exceeding $\SI{50}{\micro \metre}$ for the radiation reaction force to be greater than the perpendicular Lorentz force. Such bubble sizes are not currently realizable in current experiments. As electron beams reach higher energies the effects described here will become more significant, for example electron beams with energies in the order of 100 GeV, would need to be smaller than $8\, \mu$m to be in the slow cooling regime, where the accelerating potential within a bubble structure can sustain acceleration while radiating.

This results in an initial overdamping of the betatron oscillations. This overdamping rapidly stops as the amplitude is suddenly reduced within less than a betatron oscillation and then enters the slow damping regime. This is seen in Fig. (\ref{fig:trajectories_ion}), where both regimes are shown. The darker regions of the streamlines, show regions where no perpendicular damping occurs, \emph{i.e.}, the center of the oscillation where there is no focusing $\mathbf{E}$ field and the highest point in $r_\perp$ as at that point the perpendicular velocity is $0$. Left demonstrates the trajectories of slow cooling particles, where the betatron orbits in phase-space slowly collapse in radius, and still maintain betatron oscillation. In contrast, the right panel of Fig. (\ref{fig:trajectories_ion}) shows how particles with large amplitudes converge and collapse to regions of lower amplitude, before completing a full betatron orbit, until the betatron oscillations are restored. This happens within the region enclosed in the dashed black circle, which \reviewthree{corresponds} to $\reviewthree{A_\perp}<A_b=\sqrt{3\sqrt{2} }/(\reviewthree{r_e^{1/2}} \gamma^{3/4})$. This collapse occurs when particles are falling back into the spatial \reviewthree{equilibrium} point of oscillations, under strong fields the particles radiate more than they gain as they fall in the potential and collapse the phase-space.

This regime rapidly establishes phase-space bunching, as a significant portion of phase-space collapses. Then the amplitude is reduced enough that particles with small amplitude perform regular betatron oscillations. In contrast, the slow cooling regime, more relevant for current laboratory platforms, necessitates several betatron oscillations to establish any phase-space bunching. The fast cooling regime will be further explored in the future.
We conjecture that experiments operating in high-density ($\gtrsim10^{20}$ cm$^{-3}$) and high-energy beams ($\gtrsim 10$ GeV) will access the fast cooling regime, further enhancing bunching. Current experimental platforms cannot drive blowouts large enough or employ energetic enough particle beams to reach the fast cooling regime.

\section{Estimate for $L_d/L_r$}

We estimate the ratio between the driver depletion length $L_d$ and the ring formation length $L_r$, allowing the driver and the trailing (witness) beam to have distinct Lorentz factors $\gamma_d$ and $\gamma_w$.

We start from the estimate for the ring formation length given in the main text,
\begin{equation}
L_r \simeq L_{r,0}\,
\left(\frac{n_0}{10^{19}\,\mathrm{cm}^{-3}}\right)^{-2}
\left(\frac{E_w}{10\,\mathrm{GeV}}\right)^{-1}
\left(\frac{\sigma_\perp}{5\,\mu\mathrm{m}}\right)^{-2},
\end{equation}
where \reviewthree{$L_{r,0}\simeq 4.6\,\mathrm{cm}$ is the ring-formation length $ct_r$ with $t_r=8/(r_e\gamma_0\sigma_\perp^2)$ evaluated at the reference parameters; strong acceleration shortens it slightly, to $\simeq4.2$\,cm for $E=100$\,GV/m}, $E_w$ is the witness-beam energy, and $\sigma_\perp$ is the transverse beam waist.

To estimate the driver depletion length $L_d$ in the beam-driven blowout regime, we balance the driver energy per unit length with the wake power extracted per unit length. The driver energy per unit length is
\begin{equation}
\mathcal{E}_{\rm driver/length}\simeq n_b\,\pi\sigma_\perp^2\,\gamma_d m_e c^2,
\end{equation}
while the wake power per unit length can be written as $P_{\rm wake}\simeq E_{\rm wake} I$, with beam current
\begin{equation}
I\simeq e c\,n_b\,\pi\sigma_\perp^2.
\end{equation}
In the blowout regime, the accelerating field scales as
\begin{equation}
E_{\rm wake}\sim \left(\frac{n_b}{n_0}\right)^{1/2}\frac{m_e c\omega_p}{e},
\end{equation}
up to order-unity geometric factors \cite{lu2006nonlinear,lu2007generating}. Combining these expressions and cancelling common factors yields
\begin{equation}
L_d \simeq \frac{\mathcal{E}_{\rm driver/length}}{P_{\rm wake}}
\simeq \frac{\gamma_d\,c}{\omega_p}\left(\frac{n_b}{n_0}\right)^{-1/2},
\end{equation}
where $\omega_p=\sqrt{4\pi n_0 e^2/m_e}$ is the plasma frequency.

Combining $L_d$ and $L_r$, we obtain
\begin{equation}
\frac{L_d}{L_r}
\simeq
\gamma_d\,
\left(\frac{n_b}{n_0}\right)^{-1/2}
\left(\frac{n_0}{10^{19}\,\mathrm{cm}^{-3}}\right)^{\reviewthree{2}}
\left(\frac{E_w}{10\,\mathrm{GeV}}\right)
\left(\frac{\sigma_\perp}{5\,\mu\mathrm{m}}\right)^{2}
\frac{c}{\omega_p L_{r,0}} .
\label{eq:ratio_prefactor}
\end{equation}
Here $L_{r,0}$ is the reference ring-formation length defined in the main text, and $\omega_p=\sqrt{4\pi n_0 e^2/m_e}$.

For the parameters considered in this work, we take a plasma density $n_0=5\times10^{19}\,\mathrm{cm}^{-3}$, a driver peak density $n_b\simeq n_0$ \reviewthree{focused to a transverse size $\sigma_\perp=5\,\mu\mathrm{m}$ (a lower driver density only lengthens $L_d$)}, and a witness beam with comparable transverse size and energy $E_w=10\,\mathrm{GeV}$. Using $\gamma_d\simeq E_d/(m_ec^2)\simeq2\times10^{4}$ and the reference ring-formation length \reviewthree{$L_{r,0}\simeq4.6\,\mathrm{cm}$, we find $L_r\simeq1.9$\,mm and $L_d\simeq15$\,mm, i.e.\ $L_d/L_r\simeq8$. This indicates that, for the parameter regime explored here, the driver depletion length exceeds the ring-formation length by about an order of magnitude, so that driver evolution does not suppress the formation of ring-bunched beams within the interaction length.}

\section{Conditions for coherent betatron emission in the ICL framework}

For completeness, we reproduce here the conditions for coherent or near-coherent radiation amplification in free-electron laser (FEL) and ion-channel laser (ICL) configurations following Ref.~\cite{davoine2018ion}. In such schemes, collective emission requires both a sufficiently small relative energy spread and a sufficiently narrow distribution of the wiggler parameter $K$,
\begin{equation}
    \frac{\Delta \gamma}{\gamma} < \frac{2}{3} \rho,
\end{equation}
and
\begin{equation}
    \frac{\Delta K}{K} < \frac{2+K^2}{2K^2}\rho,
\end{equation}
where $\gamma$ is the Lorentz factor of the electrons and $\rho$ is the Pierce parameter \cite{chen1991unified, whittum1990ion},
\begin{equation}
    \rho = \left[\frac{I}{I_A}
    \frac{2(2+K^2)^2 [\mathrm{JJ}]^2}{(4+K^2)^2 \gamma}\right]^{1/3}.
\end{equation}
Here $I$ is the beam current, $I_A$ is the Alfvén current, and
\(
[\mathrm{JJ}] = J_0\!\left(\frac{K^2}{4+2K^2}\right)
- J_1\!\left(\frac{K^2}{4+K^2}\right),
\)
with $J_n$ the $n$th-order Bessel function.

The factor $2(2+K^2)^2[\mathrm{JJ}]^2/(4+K^2)^2$ is bounded between $1/2$ and $0.97$, with the most restrictive condition obtained in the small-$K$ limit, yielding \cite{davoine2018ion}
\begin{equation}
    \rho = \left[\frac{1}{2}\frac{I}{I_A}\frac{1}{\gamma}\right]^{1/3}.
\end{equation}
A sufficient condition on the relative energy spread then follows as
\begin{equation}
    \frac{\Delta \gamma}{\gamma} < \frac{2}{3}\rho .
\end{equation}

Similarly, a sufficient condition on the relative spread of the wiggler parameter reduces to
\begin{equation}
    \frac{\Delta K}{K} < \rho' \equiv 2^{-2/3}\rho .
\end{equation}

\subsection{Evolution of the wiggler-parameter distribution under betatron cooling}

In our notation, the wiggler parameter of an individual electron is proportional to its betatron amplitude, i.e.\ to $A$ at fixed $\gamma$\reviewthree{, with $A$ and $\tau$ as defined in Eq.~\eqref{eq:closed} and $\sigma_\perp$ the corresponding initial width, $\sigma_\perp=\gamma_0^{1/4}\sigma_{\perp,\rm phys}$}. The mean wiggler parameter and its spread can therefore be computed from the distribution $f(A,t)$ as
\begin{equation}
    K = \int_0^{\infty} A^2 f(A,t)\,2\pi\,dA,
\end{equation}
and
\begin{equation}
    \Delta K^2 = \int_0^{\infty} (A-K)^2 f(A,t)\,2\pi A\,dA .
\end{equation}

For an initially Gaussian (emittance-matched) beam,
\(
f_0(A) = e^{-A^2/(2\sigma_\perp^2)}/(2\pi\sigma_\perp^2),
\)
where $\sigma_\perp$ is the initial transverse beam size, one finds
\begin{equation}
    K = \frac{\sqrt{\pi}}{2\sqrt{\tau}}
    U\!\left(\frac{1}{2},0,\frac{1}{2\sigma_\perp^2\tau}\right),
    \label{app:k}
\end{equation}
where $U(a,b,c)$ is the confluent hypergeometric function of the second kind \cite{abramowitz1965handbook}. In the asymptotic limit $\tau\to\infty$,
\(
U(1/2,0,1/2\sigma_\perp^2\tau)\to 2/\sqrt{\pi},
\)
and therefore $K\sim\tau^{-1/2}$.

The corresponding spread is
\begin{equation}
    \Delta K =
    \sqrt{
    \frac{1}{\tau}
    -\frac{\pi}{4\tau}
    U\!\left(\frac{1}{2},0,\frac{1}{2\sigma_\perp^2\tau}\right)^2
    -\frac{e^{-2\sigma_\perp^2\tau}}{2\sigma_\perp^2\tau^2}
    \Gamma\!\left(0,\frac{1}{2\sigma_\perp^2\tau}\right)
    },
    \label{app:deltak}
\end{equation}
where $\Gamma(a,b)$ is the incomplete gamma function. For any fixed $\sigma_\perp$, one finds $\Delta K/K \to 0$ as $\tau$ increases, indicating that betatron cooling drives a progressive narrowing of the wiggler-parameter distribution.

To obtain a compact analytical estimate, we approximate $K\simeq r_r(\tau)$, where $r_r$ is the phase-space ring radius defined in Eq.~(6) of the main text, and $\Delta K \simeq 2(\tau^{-1/2}-r_r)$. Numerical evaluation of Eqs.~\eqref{app:k} and \eqref{app:deltak} shows that this approximation slightly overestimates $\Delta K/K$. The characteristic time at which the sufficient wiggler-spread condition $\Delta K/K<\rho'$ is approached is therefore estimated as
\begin{equation}
    \tau_c >
    \frac{(2+\rho')^2}{4\rho'(4+\rho')\sigma_\perp^2}.
\end{equation}

For experimentally accessible parameters, \emph{e.g.}, plasma density $n_0=5\times10^{19}\,\mathrm{cm^{-3}}$, beam charge $q_b\sim1$~nC, beam length $\sigma_z\sim5\,\mu$m, transverse beam size $\sigma_\perp=5\,\mu$m, and energy $E\sim10$~GeV, we estimate $\rho\sim0.05$, corresponding to a characteristic propagation length \reviewthree{$ct_c\simeq3$~cm, using $\tau\simeq(r_e/16)\gamma_0^{1/2}t$ and $\sigma_\perp=\gamma_0^{1/4}\sigma_{\perp,\rm phys}$}.

Finally, for completeness, we recall that the radiation power gain length in the ICL framework is given by \cite{rosenzweig1997space,davoine2018ion}
\begin{equation}
    L_{\mathrm{GP}} =
    \frac{2(2+K^2)}{(4+K^2)\sqrt{3}\rho}\,
    \frac{c}{\omega_\beta},
\end{equation}
which must remain shorter than the inverse beam plasma wavenumber $k_{\mathrm{pb}}^{-1}$ to avoid damping. For the parameters above, this yields an order-of-magnitude estimate $L_G\simeq 2/(\sqrt{3}\rho)\,c/\omega_\beta\sim3$~mm.

The transition to the undulator regime ($K<1$) occurs when the ring radius matches the betatron wavelength, i.e.,
\begin{equation}
r_r(t_\mathrm{u}) = \sqrt{\frac{2}{\gamma}}\,\frac{c}{\omega_p}.
\end{equation}
Recalling that \reviewthree{$r_r=\tau^{-1/2}$ in the scaled amplitude $A$, i.e.\ a ring radius $\gamma^{-1/4}\tau^{-1/2}$ in $A_\perp$, with $\tau\simeq(r_e/16)\gamma^{1/2}t$ at late times}, the corresponding time is
\begin{equation}
\reviewthree{t_\mathrm{u} = \frac{8}{r_e}\,\omega_p^{-1},}
\end{equation}
which is independent of beam parameters. This invariance arises because increasing \(\gamma\) enhances the radiative cooling rate while simultaneously shortening the betatron wavelength by the same factor of \(\gamma^{-1/2}\), leaving \(t_\mathrm{u}\) unchanged. The associated propagation length is
\begin{equation}
c t_\mathrm{u} \simeq \reviewthree{\SI{80}{\meter}}
\left[\frac{n_e}{10^{21}\,\si{\centi\meter^{-3}}}\right]^{-1},
\end{equation}
indicating that \reviewthree{the undulator regime is not reached within a single stage at these densities and} will require staging \cite{clayton2009towards,schroeder2023linear}.

\end{widetext}

\end{document}